\documentclass[11pt,oneside]{article} 

\usepackage{a4wide}

\usepackage{titlesec}

\usepackage{amsmath}
\usepackage{color}
\usepackage{framed}
\usepackage{tikz}
\usepackage{tikz-cd}

\RequirePackage{amsmath}
\RequirePackage{amssymb}
\RequirePackage{amsthm}
\RequirePackage{color}
\RequirePackage{url}
\RequirePackage{mdwlist}

\RequirePackage[all]{xy}
\CompileMatrices
\RequirePackage{graphicx}

\usepackage{xcolor}
\usepackage{amsmath,amsfonts,amssymb}
\usepackage{graphicx}
\usepackage[caption=false]{subfig}
\usepackage{enumerate}
\usepackage{mathrsfs}

\definecolor{darkblue}{RGB}{0,0,127} 
\definecolor{darkgreen}{RGB}{0,150,0}

\usepackage[normalem]{ulem}

\usepackage{setspace}   

\newcommand{\Eref}[1]{(\ref{#1})}

\newcommand{\SRef}[1]{Section~\ref{#1}}

\def\Complex{\mathbb{C}}

\def\R{\mathbb{R}}
\def\Z{\mathbb{Z}}
\def\Ham{H}
\def\Pauli{\mathcal{P}}

\def\GL{\mathrm{GL}}

\def\Stab{S}

\newcommand{\ket}[1]{|{#1}\rangle}

\newcommand{\bra}[1]{\langle{#1}|}

\newcommand{\braket}[2]{\langle{#1}|{#2}\rangle}

\newcommand{\bket}[1]{\bigl|\,{#1}\,\bigr\rangle}
\newcommand{\bbra}[1]{\bigl\langle\,{#1}\,\bigr|}

\def\Span#1{\langle #1 \rangle}

\def\smbox#1{\ \ \mbox{#1}\ \ }

\newcommand{\Field}{\mathcal{F}}

\def\Ker{\mathrm{ker}}

\def\Fn{\Field^n}
\def\Fm{\Field^m}

\def\Fnd{\Field_{n}}
\def\Fmd{\Field_{m}}
\def\Frd{\Field_{r}}

\renewenvironment{framed}
{\begin{samepage}
\MakeFramed{\hsize0.8\linewidth\advance\hsize-\width\FrameRestore}}
{\endMakeFramed\end{samepage}}

\newcommand\dolemma[1]{\vskip 5pt \noindent{\bf \underline{Lemma #1.}\ }}
\newcommand\doproposition[1]{\vskip 5pt \noindent {\bf \underline{Proposition #1.}\ }}
\newcommand\dotheorem[1]{\vskip 5pt \noindent {\bf \underline{Theorem #1.}\ }}
\newcommand\doexample[1]{\vskip 5pt \noindent {\bf \underline{Example #1.}\ }}
\newcommand\doproof{\vskip 5pt \noindent{\bf \underline{Proof:}\ }}

\newcommand\tombstone{\rule{.36em}{2ex}\vskip 5pt}

\newcounter{numitem}
\newcommand{\numitem}[1]{\refstepcounter{numitem}\thenumitem\label{#1}}

\title{Spectra of Gauge Code Hamiltonians}

\author{Simon Burton\\
Centre for Engineered Quantum Systems,\\
School of Physics,\\
The University of Sydney}

\date{\today}

\begin{document}

\maketitle

\begin{abstract}
We study the spectral gap of frustrated spin (qubit)
Hamiltonians constructed from quantum subsystem (gauge) codes.
Such a Hamiltonian can be block diagonalized, with
blocks labelled by eigenvalues of extensively many
integrals of motion (stabilizers of the code.)
Of particular interest is the 3D gauge color code,
whose Hamiltonian has been conjectured to act as a quantum
memory at finite temperature. 
Using Perron-Frobenius theory we constrain the
location of first and second eigenvalues among the
blocks of the Hamiltonian.
This allows us to numerically find the gap of some large instances
of the 3D gauge color code, which is compared to 
other frustrated spin Hamiltonians.
Finally, we suggest a relation between bounded 
stabilizers and gapped spectra.
\end{abstract}


\tableofcontents


\section{Introduction}

In this paper we study the spectra of 
Hamiltonians built from spin one-half (qubit) Pauli operators.
Sometimes these operators lead a double life,
both as the energetic terms of a Hamiltonian model, and as
the operators in a quantum error correcting code.
In this case we have
extensively many integrals of motion.
These are the stabilizers of the code,
and can be used to block diagonalize the Hamiltonion.
This sets these models apart from other 
quantum condensed matter Hamiltonions, such as
the Ising model, which has very few integrals of motion.



In Section \ref{Sec2} we show graphical depictions of
some simple Hamiltonions and introduce the action of
the Pauli $X$ and $Z$ operators.
In Section \ref{Sec3} we study the group representation theory
of subgroups of the (real) Pauli group on $n$ qubits.
In Section \ref{Sec37} we find the 
irreducible representations of an arbitrary CSS gauge (subsystem) code.

In the computational basis, 
CSS gauge code Hamiltonians have positive off-diagonal
entries and so can be viewed as the adjacency matrix
of a weighted graph.
This motivates the ideas in Section \ref{Sec5}, where 
we use Perron-Frobenius theory to
describe the low energy spectrum of such Hamiltonians.
These are new results.
Note that in this paper we use a ``neg-Hamiltonian'' convention,
so that the groundspace corresponds to the \emph{highest} eigenvalue.

In Section \ref{Sec44} we introduce the 3D gauge color code Hamiltonian%
\cite{Bombin2015,Bombin2015single,Kubica2015}.
We find that this model 
decomposes into six mutually commuting operators in Section \ref{Sec45}.
This gives an exponential reduction in the numerical
difficulty of exactly diagonalizing this Hamiltonian,
results of which we present in Section \ref{Sec6}.
These numerics show some evidence of a spectral
gap, which is contrasted against models that are known
to be gapless: the two and three dimensional compass model,
the one dimensional $XY$-model and 
the one dimensional transverse field Ising model.
The conjecture that the gauge color code model is gapped
in turn lends weight to the possibility that this may
be a self-correcting quantum memory~\cite{Brown2016}.
Having a constant gap (bound from below)
is also part of the story of topologically ordered phases
\cite{Kitaev2003,Brown2016}.
We also apply these techniques to a model
we call the 2D $XY$-plaquette model,
which has weight four gauge operators.

Finally in Section \ref{Sec7} 
we investigate the connection between the size of
stabilizer generators
and the gapped nature of the Hamiltonian.
In particular, we suggest a relation between
extensive stabilizer generators, and gapless excitations of
the Hamiltonian.
This is further application of the Perron-Frobenius 
theory in the context of Cheeger inequalities.

\vskip 5pt \noindent{\bf \underline{Acknowledgements.}\ }
The author would like to thank 
Andrew Doherty,
Steven Flammia,
Michael Kastoryano,
Jiannis Pachos and 
Dan Browne.

\subsection{Some motivating examples}\label{Sec2}

\doexample{\numitem{Ex1}}
We start our journey considering a two-dimensional state space.
This space is blessed with two basis vectors $\ket{0}$ and $\ket{1}.$
The $Z$ and $X$ operators act on these states as:
\begin{center}
\includegraphics[]{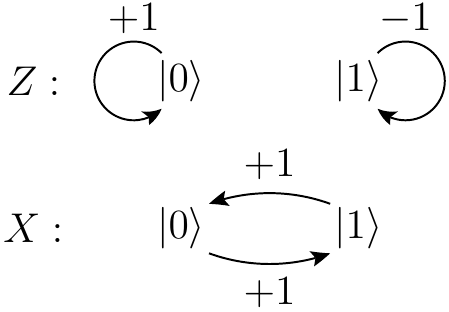}
\end{center}
From this picture we can see that $Z$ acts by \emph{stabilizing} the
state $\ket{0}$ and anti-stabilizing the $\ket{1}$ state.
The $Z$ operator has been reduced 
to two operators each acting on a one dimensional subspace:
$Z = +1 \oplus -1.$
The $X$ operator serves to ``bitflip'' the state between
these two subspaces.

But what happens if we get confused and end up swapping
the $X$ and $Z$ operators? We would like to see the $X$ operator
as stabilizing / anti-stabilizing two subspaces, together with the
$Z$ operator as bitflipping between these.
The trick is to consider the \emph{orbits} of the operator
we hope to act as a stabilizer.
In this case there is only one orbit, $\ket{0}+\ket{1}$
and indeed, the $Z$ operator bitflips this to another state
$\ket{0}-\ket{1}$ that is anti-stabilized by $X.$
\tombstone

We are going to be considering Hamiltonians built from
summing operators of this form.
In this paper we use a ``neg-Hamiltonian'' convention,
to save complicating expressions with negative signs.
The ground space corresponds to the \emph{highest} eigenvalue.

So building a Hamiltonian from a single $X$ or $Z$ term,
we find the ground space as the stabilized space
by summing over the orbit of that term.
The other operator,
which we call the \emph{adjacent operator}, acts to bitflip
between the eigenspaces.

\doexample{\numitem{Ex2}}
To further elucidate this idea we turn to another example,
which is a Hamiltonian built from three commuting and
independent operators:
$$
    \Ham = XXI + IXX + ZZZ.
$$
Starting with $\ket{000}$
the terms of the Hamiltonian generate an orbit
given by $$\mathrm{Orbit}(\ket{000}) = \{\ket{000}, \ket{011}, \ket{110}, \ket{101}\}.$$
Notice that the $ZZZ$ term fixes all these states.
Summing over this orbit we get the following ground state:
$$\mathrm{GS} = \ket{000}+\ket{011}+\ket{110}+\ket{101}.$$

We select three adjacent operators
$ZII, IIZ,$ and $IXI,$
one for each of the stabilizer operators:
\begin{align*}
    ZII : \mbox{GS} &\mapsto \ket{000}+\ket{011}-\ket{110}-\ket{101} 
        \ \mbox{anti-stabilized by}\  XXI,\\
    IIZ : \mbox{GS} &\mapsto \ket{000}-\ket{011}+\ket{110}-\ket{101} 
        \ \mbox{anti-stabilized by}\  IXX,\\
    IXI : \mbox{GS} &\mapsto \ket{010}+\ket{001}+\ket{100}+\ket{111} 
        \ \mbox{anti-stabilized by}\  ZZZ.
\end{align*}

These adjacent operators form an abelian group
of order $2^3 = 8$ and by applying each element of
this group to the ground state we get a basis
of our state space, which we call a \emph{symmetry
invariant basis.}

The adjacent operators are not unique in general. 
For this example we could have also chosen $IZZ, ZZI, XXX.$
\tombstone

\doexample{\numitem{Ex3}}
Now we consider a four qubit example:
$$
    \Ham = XXII + IIXX + ZIZI + IZIZ.
$$
This time the terms of the Hamiltonian do not generate 
an abelian group.
We will call this group, as generated by the terms in the Hamiltonian,
the \emph{gauge group}, $G$.
The \emph{stabilizer} subgroup of $G$ will be the elements of $G$
that commute with every other element in $G.$
By inspection we see this group is generated by $\Stab_0=\{XXXX, ZZZZ\}.$
We can extend these generators to a complete independent generating
set for $G$ using the operators  $R_0=\{XXII, ZIZI\}.$
These $R_0$ operators generate the \emph{reduced gauge group} $R.$
The operators adjacent to $\Stab_0$ we
call the \emph{error operators} $T_0$. 
We choose $T_0 = \{ZZZI, IIIX\}.$
(Once again, these are not unique.)
The \emph{logical operators} are the $n$-qubit 
Pauli operators outside of the group
$G$ that commute with $G.$
In this case they are generated by $L_0 = \{XIXI, ZZII\}.$
All of this can be summarized in a table of adjacent pairs:
\begin{center}
\includegraphics[]{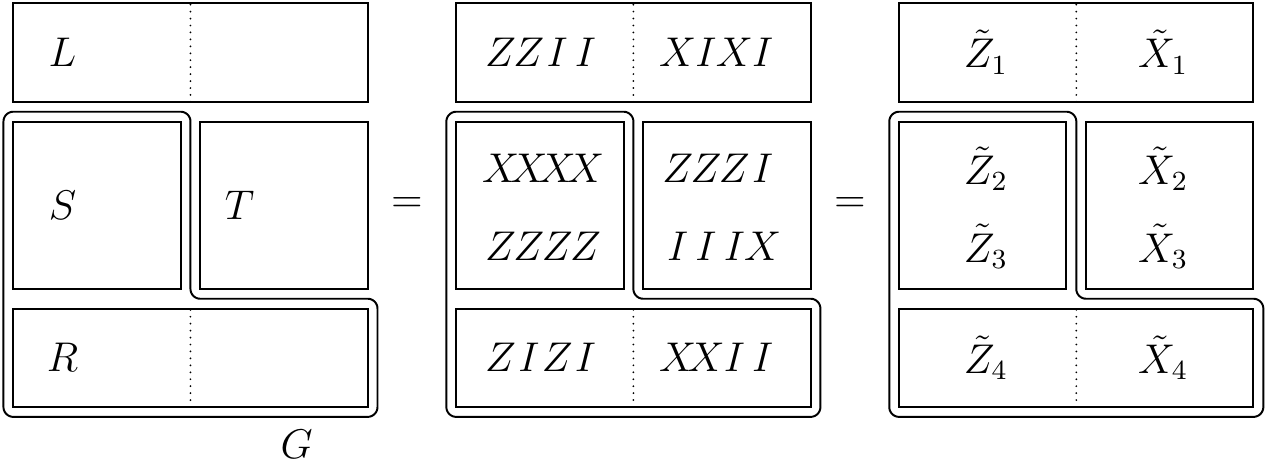}
\end{center}
where the number of rows equals $n,$ 
each operator commutes with operators on other rows,
and anticommutes with the operator on the same row. 
There is no need to include any phases ($\pm I$) in these tables
because phases commute with everything.
If we take all the entries in the left column
we get the operators 
$\{ ZZII, XXXX, ZZZZ, ZIZI \}.$ 
These generate an abelian group 
that stabilizes the
state $\ket{\psi} = \ket{0000}+\ket{1111}.$
Let $r$ be the gauge operator $XXII$ adjacent to the 
stabilizer $ZIZI$,
The state $\ket{\psi}$ then lies in the $G-$orbit 
$$
\{\ket{\psi}, r\ket{\psi}\} = \{\ket{0000}+\ket{1111}, \ket{1100}+\ket{0011}\}.
$$
We use the $T_0$ operators $t_1=ZZZI$ and $t_2=IIIX$
to list three other $G-$orbits:
$$
\{t_1 \ket{\psi}, t_1 r\ket{\psi}\}, 
\{t_2 \ket{\psi}, t_2 r\ket{\psi}\}, 
\{t_1 t_2 \ket{\psi}, t_1 t_2 r\ket{\psi}\}.
$$
So now we have sixteen vectors, forming an orthogonal basis for the state space.
This is the symmetry invariant basis for this Hamiltonian.

We can now arrange these basis vectors on the vertices of a four dimensional hypercube,
such that each dimension corresponds to one of the adjacent $\tilde{X}_i$ bitflip
operators.
Such an arrangement has a cartesian product
structure which induces a tensor product decomposition of
the original state space that corresponds to the $\tilde{X}_i:$
\begin{center}
\includegraphics[width=0.7\columnwidth]{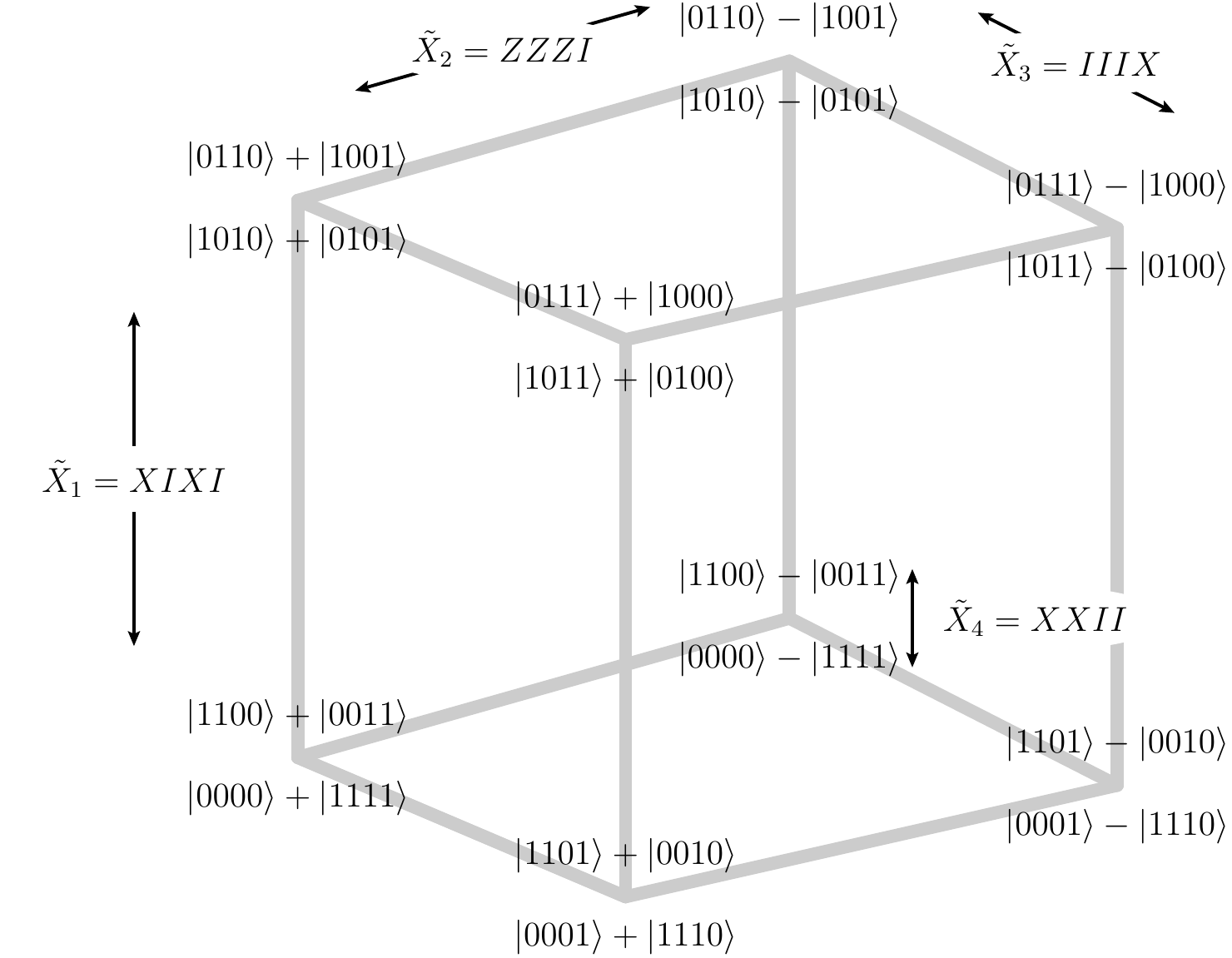}
\end{center}
The Hamiltonian acts on states by left multiplication.
Because this action
is a sum of gauge group elements,
it will decompose into blocks,
one for each $G-$orbit.
We depict this action as a weighted graph, where we omit edges
with zero weight:
\begin{center}
\includegraphics[width=0.7\columnwidth]{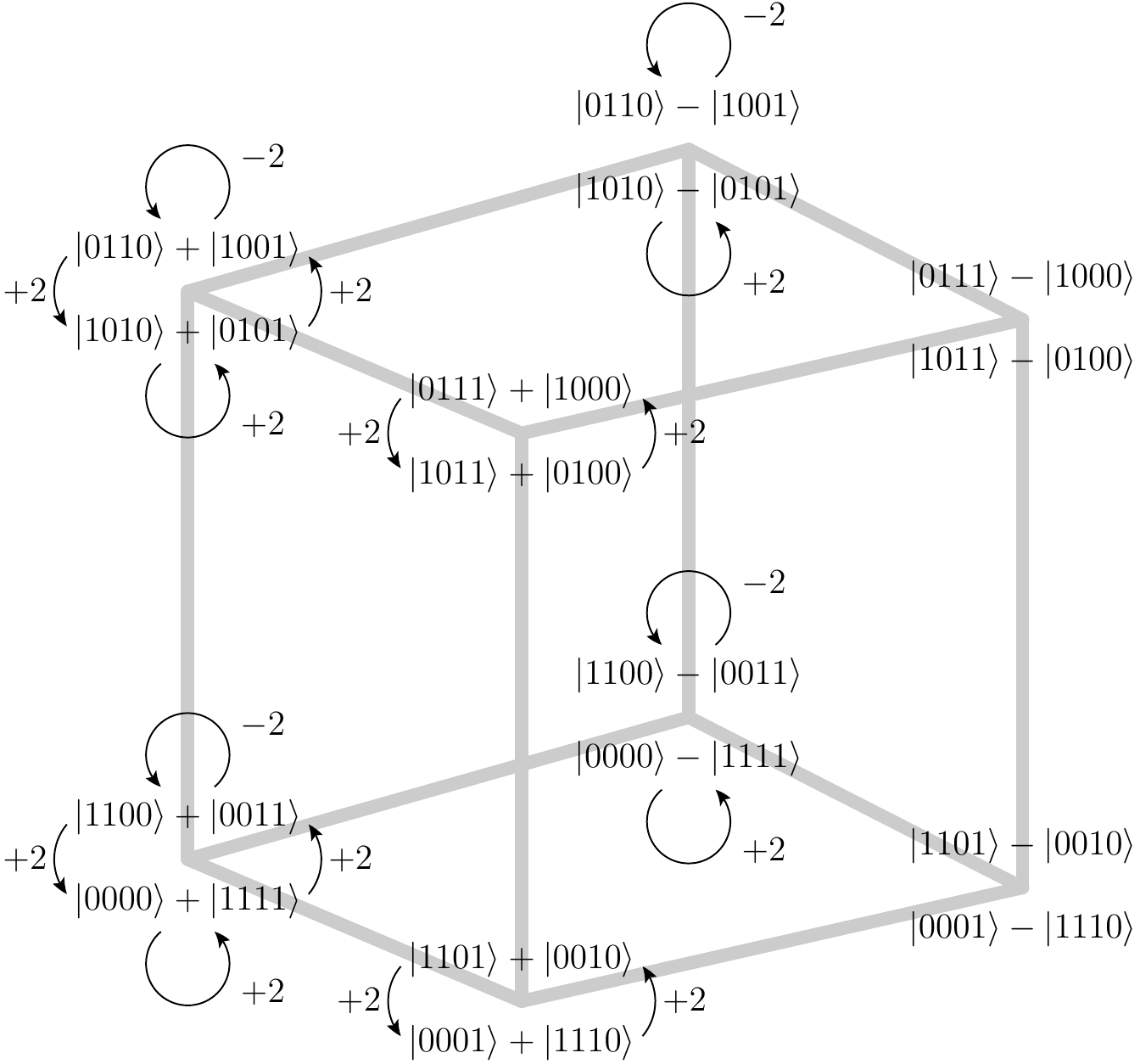}
\end{center}

\def\Xt{\tilde{X}}
\def\Zt{\tilde{Z}}

Equivalently we use this basis to 
write the matrix for the Hamiltonian in
block diagonal form:
$$
\Ham = 
\left( \begin{array}{cccc}
2(X+Z) & 0 & 0 & 0 \\
0  & 2X & 0 & 0 \\
0  & 0 & 2Z & 0 \\
0  & 0 & 0 & 0 \\
\end{array} \right) \otimes I
$$
This block diagonal form will be worked out for
a general gauge group below and summarized in Section \ref{Sec35}.
\tombstone

So far we have been analysing the Hamiltonian in the
symmetry invariant basis. The complete story of how this
works is in \SRef{Sec3} below.
But we can also just fix the computational basis and
examine the Hamiltonian in this basis.
This leads to the Perron-Frobenius theory in \SRef{Sec5}.


\section{Gauge code Hamiltonians}\label{Sec3}

\subsection{The Pauli group}\label{Sec31}

The Pauli group $\Pauli_1$ is normally 
defined as a set of matrices closed under
matrix multiplication,\footnote{The original definition of the Pauli
group also includes an imaginary unit $i$, which we
do not include. So perhaps this should be called the
\emph{real} Pauli group.}
but we can define it abstractly
as the group generated
by the (abstract) elements $\{\omega, X, Z\}$ with
relations as follows:
$$
\omega^2=I,\ X^2=I,\ Z^2=I,\ \omega X\omega X=I,\ \omega Z\omega Z=I,\ \mbox{and}\  \omega ZXZX=I,
$$
where $I$ is the group identity.
Actually, $\omega $ is generated by $X$ and $Z$,
so it is not necessary to include $\omega $ in the generating set,
but here it simplifies the relations.
This group has eight elements, and is isomorphic to the dihedral group $D_4$,
the symmetry group of a square.

To define
the {\it $n$-qubit Pauli group} $\Pauli_n$, 
we use the $2n+1$ element 
generating set 
$$\{\omega , X_1, .., X_n, Z_1, .., Z_n\}$$
with relation $\omega^2=I$ as before, and
\begin{equation}\label{presentation}
\begin{array}{c}
X_i^2=I,\ Z_i^2=I,\ \omega X_i\omega X_i=I,\ \omega Z_i\omega Z_i=I,\ \omega Z_iX_iZ_iX_i=I, 
\mbox{\ for\ } i=1,...n,\\
X_iX_jX_iX_j=I,\ 
Z_iZ_jZ_iZ_j=I,\ 
Z_iX_jZ_iX_j=I, \mbox{\ for\ } i, j = 1,..,n,\ i\ne j.
\end{array}
\end{equation}

This abstract approach to the definition of a group is known as
a group \emph{presentation}. In general, this is a set of
generators together with a set of relations satisfied
by these generators.

Note that each of the generators squares to the identity,
and of these, only $\omega$ commutes with every element of $\Pauli_n.$
Therefore we will write $\omega$ as $-I,$
similarly $\pm I$ will denote the
set $\{\omega, I\},$ and $-X$ is $\omega X$, etc.

We write the group commutator as
$[[g, h]]:=ghg^{-1}h^{-1}$
and note the important commutation relation:
$$
    [[Z_i, X_j]] = 
    \left\{ \begin{array}{ll}
 -I &\mbox{if}\ i=j,\\
 I &\mbox{if}\ i\ne j.\end{array}\right.
$$
If we take an arbitrary $g\in \Pauli_n$
written as a product of the generators,
it follows that we can rewrite this
product uniquely as 
$ g = \pm g_1 ... g_n $
where each $g_i$ is one of $I, Z_i, X_i$ or $X_i Z_i$
for $i=1,..,n.$
Therefore, the size of the
Pauli group is 
$$
    |\Pauli_n| = 2^{2n+1}.
$$

The subgroup of $\Pauli_n$ generated by
the elements $\{X_1,...,X_n\}$ 
is denoted $\Pauli_n^X.$ These are the $X$-type
elements. Similarly,
 $\{Z_1,...,Z_n\}$ generates 
the subgroup of $Z$-type elements $\Pauli_n^Z$.

We now define the
{\it Pauli representation} 
of the Pauli group as a group homomorphism:
$$
    \rho_{\mathrm{pauli}} : \Pauli_n \to \GL(\Complex[2^n])
$$
where $\Complex[2^n]$ is the $2^n$-dimensional state space of $n$ qubits.
On the independent generators 
$\{X_1, .., X_n, Z_1, .., Z_n\},\ \rho_{\mathrm{pauli}}$
is defined as the following tensor product of $2\times 2$ matrices:

\begin{align*}
\rho_{\mathrm{pauli}}(X_i) := &\bigotimes_{j=1}^n \left\{ \begin{array}{ll}
\left( \begin{array}{ll}
0&1\\
1&0\end{array} \right) &\mbox{for $j=i$}\\
\\
\left( \begin{array}{ll}
1&0\\
0&1\end{array} \right) &\mbox{for $j\ne i$} \end{array}
\right.\\
\rho_{\mathrm{pauli}}(Z_i) := &\bigotimes_{j=1}^n \left\{ \begin{array}{ll}
\left( \begin{array}{ll}
1&0\\
0&-1\end{array} \right) &\mbox{for $j=i$}\\
\\
\left( \begin{array}{rr}
1&0\\
0&1\end{array} \right) &\mbox{for $j\ne i$}\end{array}
\right.
\end{align*}

\subsection{Subgroups of the Pauli group}\label{Sec32}

We now define an {\it $n-$qubit gauge group} to be 
any non-abelian subgroup $G$ of $\Pauli_n,$
defined by a set of generators $G_0\subset \Pauli_n,$
$$ G := \langle G_0\rangle.$$
Because $G$ is not abelian, it follows that $-I\in G.$
We also restrict $G_0$ to only contain Hermitian operators,
which is equivalent to requiring that $g^2=I$ for all $g\in G_0.$

Now let $\Stab$ be the largest subgroup of $G$ not containing
$-I.$
$\Stab$ is then an abelian subgroup,
also known as the {\it stabilizer} subgroup.
$G$ decomposes as a direct product:
$$G = \Stab\times R,$$
where $R\cong \Pauli_r$ for some $1\le r\le n,$
and $\Stab\cong \Z_2^{m}$ for $0\le m<n.$
Therefore, 
$$|G| = |\Stab| |R| = 2^{m+2r+1}.$$
We call $R$ the {\it reduced gauge group}.
We consider both $\Stab$ and $R$ to be subgroups of $G.$
Let $\phi:R\to \Pauli_r$ be a group isomorphism,
then $R_0 := \{\phi^{-1}(X_i), \phi^{-1}(Z_i)\}_{i=1,..,r}$
is a set of independent generators of $R.$
We also let $\Stab_0$ be a set of $m$ independent generators of $\Stab.$

To find the cosets of $G$ in $\Pauli_n$ we take
the group closure of $\Pauli_n-G$; when this is non-empty
we only need to add $I$ and $-I.$
This is another
gauge group, whose reduced gauge group is known as
the {\it logical} operators $L$, and whose 
stabilizer subgroup is known as the {\it error} operators $T.$
Now any coset of $G$ can be written as $ltG$ with
$l\in L$ and $t\in T.$
The size of $T$ equals the size of $\Stab$: $|T|=|\Stab|=2^m.$
If we let $L_0$ be an independent generating set for $L$
then we have the important formula:
\begin{align*}
n &= \frac{1}{2}|L_0| + |\Stab_0| + \frac{1}{2}|R_0|\\
  &= k + m + r
\end{align*}
We summarize the information in this section in a table
of Pauli group elements arranged in
two columns and $n$ rows:
\begin{center}
\includegraphics[]{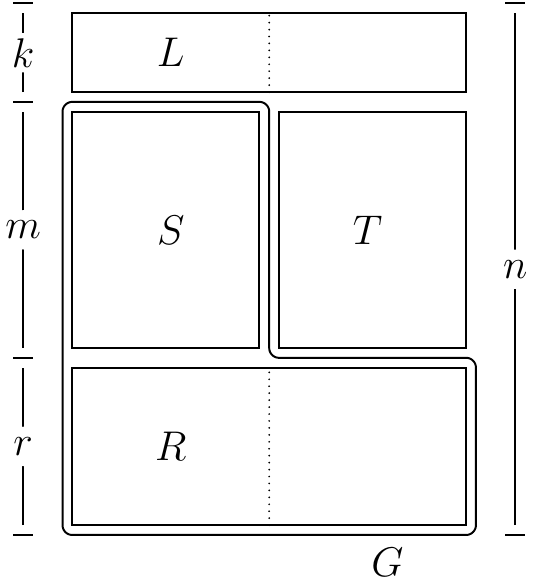}
\end{center}
Here we show the $2n$ generators of $\Pauli_n$ arranged 
so that each row contains a pair of generators,
where each such generator anti-commutes with the operator on the same row and
commutes with all the other operators in the table.
Note that this is exactly the definition of the Pauli group
via a presentation given in the previous section.
Furthermore, the table shows $2k$ generators
of $L$, $m$ generators each for $S$ and $T,$ and $2r$
generators of $R.$
The gauge group $G$ encloses $R$ and $S$, and one can
immediately see how $L$ and $T$ also form a gauge group.


\subsection{Representations of the Pauli group}\label{Sec33}

We now define the
{\it Pauli representation} 
of the Pauli group as a group homomorphism:
$$
    \rho_{\mathrm{pauli}} : \Pauli_n \to \GL(\Complex[2^n])
$$
where $\Complex[2^n]$ is the $2^n$-dimensional state space of $n$ qubits.
On the independent generators 
$\{X_1, .., X_n, Z_1, .., Z_n\},\ \rho_{\mathrm{pauli}}$
is defined as the following tensor product of $2\times 2$ matrices:

\begin{align*}
\rho_{\mathrm{pauli}}(X_i) := &\bigotimes_{j=1}^n \left\{ \begin{array}{ll}
\left( \begin{array}{ll}
0&1\\
1&0\end{array} \right) &\mbox{for $j=i$}\\
\\
\left( \begin{array}{ll}
1&0\\
0&1\end{array} \right) &\mbox{for $j\ne i$} \end{array}
\right.\\
\rho_{\mathrm{pauli}}(Z_i) := &\bigotimes_{j=1}^n \left\{ \begin{array}{ll}
\left( \begin{array}{ll}
1&0\\
0&-1\end{array} \right) &\mbox{for $j=i$}\\
\\
\left( \begin{array}{rr}
1&0\\
0&1\end{array} \right) &\mbox{for $j\ne i$}\end{array}
\right.
\end{align*}


Normally the image of 
$\rho_{\mathrm{pauli}}$ is thought of as the
Pauli group itself, and we are indeed free to think
that way because $\rho_{\mathrm{pauli}}$ is a group
isomorphism (onto its image).

Given a group representation $\rho:G\to GL(V)$
the {\it character} of $\rho$ is a function
$\chi_\rho:G\to \Complex$ given by
$$
    \chi_\rho(g) = \mbox{Tr}\ \rho(g).
$$

Given two functions $u,v : G \to \Complex$ 
we define the following inner product:
$$
    \langle u, v \rangle := \frac{1}{|G|} \sum_{g\in G} u(g) \overline{v(g)}.
$$

The character of the Pauli representation, $\chi_{{pauli}}:\Pauli_n\to\Complex$
is given by:
$$
\chi_{{pauli}}(g) = \sum_{v \in basis} \langle v | \rho_{{pauli}}(g) | v \rangle
    = \left\{ \begin{array}{ll}
 \pm 2^n &\mbox{if}\ g=\pm I\\
 0 &\mbox{otherwise}\end{array}\right.
$$

Since $|\Pauli_n|=2^{2n+1}$ it follows that
$\langle\chi_{pauli},\chi_{pauli}\rangle = 1$ and
so $\rho_{pauli}$ is an irreducible representation of $\Pauli_n.$

The only other irreps of $\Pauli_n$ are 
the $1$-dimensional irreps $\rho:\Pauli_n\to\Complex$
defined on the independent generators as:
    $$ \rho(X_i) = \pm 1,\quad \rho(Z_i) = \pm 1.$$

So we have $2^{2n}$ many $1$-dimensional irreps,
and a single $2^n$-dimensional irrep.
Summing the squares of the dimensions
shows that we have a complete set of irreps of $\Pauli_n.$

%

\subsection{Representations of gauge groups}\label{Sec34}

Although $\rho_{\mathrm{pauli}}$
restricted to a gauge group $G\subset\Pauli_n$ serves as a representation
of $G$ it is no longer irreducible.
Our aim will be to decompose $\rho_{\mathrm{pauli}}$ into irreps of $G.$

The $1$-dimensional irreps $\rho:G\to \Complex,$
are now defined by
specifying the action of $\rho$ on the independent generators:
$$
    \rho(h)=\pm 1\ \mbox{for}\ h\in \Stab_0,
    \quad \rho(\phi^{-1}(X_i)) = \pm 1,\quad \rho(\phi^{-1}(Z_i)) = \pm 1.
$$
This gives all $2^{m+2r}$ of the $1$-dimensional irreps.

The $2^r$-dimensional irreps are given by:
$$
    \rho(h) = \pm I^{\otimes r}\ \mbox{for}\ h\in \Stab_0,
    \quad \rho(\phi^{-1}(X_i)) = X_i,\quad \rho(\phi^{-1}(Z_i)) = Z_i.
$$
We are free to choose the signs of the $\rho(h)$ for each $h\in \Stab_0.$
Hence there are $2^m$ many of these irreps.
Each such choice corresponds to the choice of a {\it syndrome} vector $s(h)=\pm 1$, for $h \in \Stab_0,$
or alternatively, choice of an element $t\in T:$
$$
    \rho^1_t(h) = \left\{ \begin{array}{ll}
 I^{\otimes r}\ &\mbox{if $th=ht$}\\
 -I^{\otimes r}\ &\mbox{if $th=-ht$}\end{array} \right. 
$$


Because $G$ decomposes 
into a direct product $G=\Stab\times \Pauli_r$ we have the
following representations:
$$
    \rho_t(g) = \rho^1_t(h) \rho^r_{pauli}(g'),
$$
where $g=hg'$, $h\in \Stab$, $g'\in \Pauli_r$ 
and $\rho^r_{pauli}$ is the $r$-qubit Pauli representation.
The character for this representation is:
$$
\chi_{t}(hg') = \rho_t^1(h) \sum_{v \in basis} \langle v | \rho^r_{{pauli}}(g') | v \rangle
    = \left\{ \begin{array}{ll}
 \pm 2^r\rho_t^1(h) &\mbox{if}\ g'=\pm I\\
 0 &\mbox{otherwise}\end{array}\right.
$$

We have that $|G|=2^{2r+m+1}$ and so
$\langle\chi_{t},\chi_{t}\rangle = 1$ and
$\rho_t$ is an irreducible representation of $G.$
We now count the occurrences of 
this representation in $\rho^r_{pauli}$:
\begin{align*}
\langle\chi^r_{pauli},\chi_{t}\rangle &= \frac{1}{|G|}\sum_{g\in G} \chi^r_{pauli}(g)\overline{\chi_{t}(g)} \\
&= \frac{1}{2^{2r+m+1}} \sum_{g=\pm I} 2^n 2^r = \frac{2^{n+1+r}}{2^{2r+m+1}} = 2^k
\end{align*}
where $k$ is the number of logical qubits so that $n=r+m+k.$

In summary, the Pauli representation decomposes into 
$2^m$ many irreps $\rho_t,$ 
each with dimension $2^r,$ 
and appearing with multiplicity $2^k:$
$$
    \rho_{\mathrm{pauli}} = 
        \bigoplus_{t\in T}\ \rho_t \otimes I^{\otimes k}
$$

%
%

\subsection{Block diagonalizing the Hamiltonian}\label{Sec35}

The Hamiltonian of interest is 
an operator $\Ham:\Complex[2^n]\to\Complex[2^n]$:
$$ \Ham := \sum_{g\in G_0} \rho_{\mathrm{pauli}}(g).$$
Using the above decomposition we find:
\begin{align*}
    \Ham &= \sum_{g\in G_0}\ \bigoplus_{l\in L, t\in T}\ \rho_t(g)\\
         &= \bigoplus_{l\in L, t\in T} \sum_{g\in G_0}\ \rho_t(g).
\end{align*}
We will notate each block as
$\Ham_t := \sum_{g\in G_0}\rho_t(g)$
for each irrep $\rho_t$ appearing in $\Ham.$
\begin{framed}
\noindent
The Hamiltonian is block diagonalized, with blocks indexed by operators $t$ in
the abelian group $T$ and multiplicity $2^k:$
$$
    \Ham =  \bigoplus_{t\in T}\ \Ham_t \otimes I^{\otimes k}.
$$
\end{framed}

More generally, we can assign real valued weights
$J_g\in\R$
to each operator $g\in G_0,$
\begin{align*}
    \Ham = \sum_{g\in G_0} J_g \rho_{\mathrm{pauli}}(g)
            = \bigoplus_{l\in L, t\in T} \sum_{g\in G_0}\ J_g \rho_t(g).
\end{align*}
In other words, using weights does not change the block structure of $\Ham.$

In the following sections we will forget the distinction 
between $g$ and $\rho_{\mathrm{pauli}}(g)$,
so terms such as $Z$ and $X$ can be understood
as the corresponding Pauli linear operators.

%
%

%
%

%
\subsection{Symplectic decomposition of CSS gauge codes}\label{Sec36}


In this section, and the remainder of this paper,
we restrict our
attention to gauge groups formed from terms 
in $\Pauli_n^X\cup\Pauli_n^Z.$
We call these \emph{CSS gauge codes.}
We next turn to a discussion of the symplectic structure of
these operators.

Let $\Field$ denote the finite field with two elements $0$ and $1$.
Both $\Pauli^X_n$ and $\Pauli^Z_n$ are abelian groups,
and can be identified with the additive 
group structure of the $n$ dimensional vector space
over $\Field:$
$$
    \Pauli^X_n \cong \Fn,  \ \ 
    \Pauli^Z_n \cong \Fn. 
$$
We do this in the obvious way by sending $X_i$ to the basis vector with
$1$ in the $i-$th entry, and similarly for each $Z_i$. 
We also identify the computational basis of our statespace $\Complex[2^n]$
with $\Fn$ in the obvious way:
$$
\Complex[2^n] \cong \Complex[\Fn].
$$
This has the potential to be very confusing, and
so where appropriate we use $X$ and $Z$ subscripts.

$X$-type operators act on the $\Complex[\Fn]$
basis vectors using $\Field$ addition:
$$
    g_X \in \Pauli^X_n \cong \Fn, \ \ g_X : v \longmapsto g_X + v
$$

$Z$-type operators act on the
$\Complex[\Fn]$ basis vectors using $\Field$ inner product:
$$
    g_Z \in \Pauli^Z_n \cong \Fn, \ \ g_Z : v^\top \longmapsto g_Z v^\top
$$
This is an $\Field$ scalar, just zero or one. We think of this
as a ``syndrome''.
This suggests that actually these $Z$-type operators
live in the dual vector space $\Fnd.$
Because of the underlying symmetry
(and notational confusion)
between the $X$ and $Z$-type operators,
we make the convention that by default
all our $\Field$-vectors come as row vectors (ie. dual vectors).
This means we use the transpose operator $^\top$ to
indicate a primal (column) vector.

It doesn't make sense to add an $X$-type operator and
a $Z$-type operator:
$$
    g_Z + g_X \ \ \ \mbox{\emph{don't do this!!!}}
$$
but it does make sense to take the inner product:
$$
    g_Z g_X^\top = g_X g_Z^\top.
$$
This is an $\Field$ scalar which gives the commutator of the 
two operators.

An $\Field$-linear operator such as
$ A : \Fn \to \Fm $
acts on the left as $ u^\top \mapsto A u^\top.$
It also acts on dual vectors as
$ A : \Fmd \to \Fnd $
which corresponds to acting on the right: $v \mapsto vA.$ 
We call the rowspace of 
$A$ the \emph{span} and denote 
it as 
$$\Span{A} = \{ vA | v \in \Fmd \}$$
The kernel of $A$ is defined as
$$
    \Ker(A) = \{ u^\top | u^\top \in \Fn,\ \  A u^\top = 0 \}.
$$

We wish to use this language to decompose a CSS gauge group $G.$
First we write the gauge group generators in terms of
$X$-type and $Z$-type operators:
$$
    G_0 = G_X \cup G_Z.
$$
Following the theory from the previous section,
we are going to rewrite the gauge group generators
as a union of stabilizer generators $S_0 = S_X \cup S_Z$
and reduced gauge generators $R_0 = R_X \cup R_Z.$
Similarly, the error operators
will be split into $X$-type and $Z$-type
operators $T_X$ and $T_Z$ and
finally the logical operators
$L_X$ and $L_Z.$
We summarize all of these sets
in the following table:
\begin{center}
\includegraphics[]{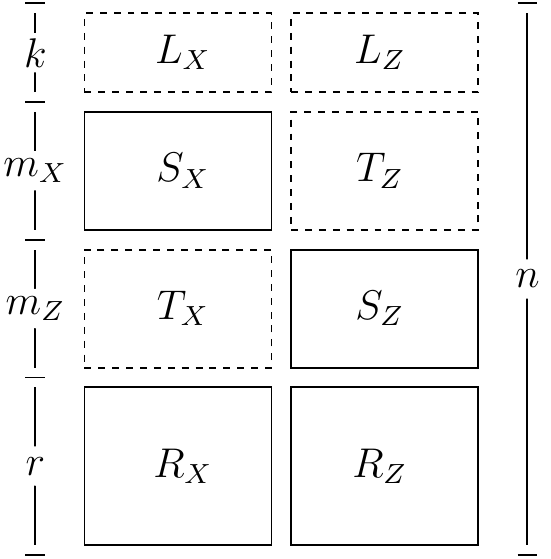}
\end{center}
The solid rectangles indicate operators that
span the $X$ and $Z$ parts of the gauge group,
and the dashed rectangles indicate operators that
do not live inside the gauge group.

We consider each of these blocks 
$L_X, L_Z, S_X, T_Z, T_X, S_Z, R_X, R_Z,$
as well as $G_X,G_Z,$
as either a set of $\Fnd$ vectors (the rows) or as an 
$\Field$-linear operator.
For example, we write $u\in R_X$ to mean $u$ is 
a row of the matrix $R_X$.

\dotheorem{(Decomposition)}
Let $G$ be an $n$-qubit 
CSS gauge group with generators $G_X\cup~G_Z.$
We can construct a solution to the following $\Field$-linear
block matrix equation:
\begin{equation}\label{lstr}
\left( \begin{array}{l}
L_X\\
S_X\\
T_X\\
R_X
\end{array} \right)
\left( \begin{array}{l}
L_Z\\
T_Z\\
S_Z\\
R_Z
\end{array} \right)^\top =
I,
\end{equation}
such that 
$\Span{S_X}+\Span{R_X} = \Span{G_X}$
and 
$\Span{S_Z}+\Span{R_Z} = \Span{G_Z},$
with $I$ the $n\times n$ identity matrix.

\doproof
We first find the stabilizers $S_Z$.
These are built out of $\Fnd$ vectors from the span of $G_Z$
that commute with the rows of $G_X:$
\begin{align*}
    \Span{S_Z} &= \{\  vG_Z \ |\  v G_Z G_X^\top = 0, \ \ v \in F_{|G_Z|}\ \} \\
               &= \{\  vG_Z \ |\  v^\top \in \Ker(G_X G_Z^\top)  \ \}.
\end{align*}
The generators (rows of $S_Z$) are then extracted
from this span by row reduction.
We swap the role of $X$ and $Z$ to find $S_X.$


We solve the block matrix equation \Eref{lstr},
subject to the restriction that the rows of
$R_X$ lie in the span of $G_X$,
the rows of $L_X$ do not,
and similarly for $R_Z$ and $L_Z.$
This set of 16 equations is quadratic in the unknown variables
and so it is not obvious how to proceed, but it turns out a
systematic way can be found.

We begin by finding $L_Z.$
These operators satisfy the following \emph{homology} condition:
$$
    l_Z \in L_Z \smbox{is given by} l_Z^\top \in \Ker(G_X) \smbox{mod} \Span{S_Z}.
$$
In other words, $L_Z$ is formed from a basis for the kernel of $G_X$ 
row-reduced using $S_Z.$
To be more specific 
we take any direct sum decomposition
$$\Fnd = \Span{S_Z} \oplus V$$
then the operation of mod $\Span{S_Z}$ is the
projection onto $V.$
We can explicitly write such a 
projector as the $n\times n$
matrix given by
$$
    P_Z = I + A^\top S_Z
$$
where the matrix $A$ is the $m_Z\times n$ matrix consisting of
the leading 1's in any row-reduction of $S_Z.$
We define $P_X$ similarly for the operation of mod $\Span{S_X}.$

To find $L_X$ we solve the following $\Field$-linear
system:
$$
\left( \begin{array}{l}
L_Z\\
G_Z\\
\end{array} \right)
L_X^\top = 
\left( \begin{array}{l}
I\\
0
\end{array} \right)
$$

The reduced gauge group matrix $R_X$
is found as a row-reduction of $G_X P_X.$
We cannot merely set $R_Z$ to be $G_Z P_Z$ 
because we also require $R_X R_Z^\top = I.$
Instead we define the auxiliary matrix
$\widetilde{R}_Z$ to be a row-reduction of  $G_Z P_Z.$

The error operators $T_X$ are then found as the solution
to the $\Field$-linear system:
$$
\left( \begin{array}{l}
L_Z\\
S_Z\\
\widetilde{R}_Z
\end{array} \right)
T_X^\top = 
\left( \begin{array}{l}
0\\
I\\
0
\end{array} \right)
$$
And then the operators $T_Z$ solve the $\Field$-linear system:
$$
\left( \begin{array}{l}
L_X\\
S_X\\
T_X\\
R_X
\end{array} \right)
T_Z^\top = 
\left( \begin{array}{l}
0\\
I\\
0\\
0
\end{array} \right)
$$
Finally at this point $R_Z$ is given as the solution to
$$
\left( \begin{array}{l}
L_X\\
S_X\\
T_X\\
R_X
\end{array} \right)
R_Z^\top = 
\left( \begin{array}{l}
0\\
0\\
0\\
I
\end{array} \right)
$$
Note that $R_Z$ and $\widetilde{R}_Z$ have identical span 
and so we have $R_Z T_X^\top = 0.$
\tombstone

We call this array of eight $\Field$-linear 
matrices
an $(L,S,T,R)$-decomposition of the gauge group.
In general this will not be unique for
any given gauge group.
Note that equation \Eref{lstr} is equivalent to the 
Pauli presentation \Eref{presentation}.

\subsection{CSS gauge code Hamiltonians}\label{Sec37}

The complex Hilbert state space of our
Hamiltonian has $2^n$ dimensions and we
write this space as $\Complex[2^n]$.
This notation is meant to suggest that
we are forming a $\Complex$ vector space
using $2^n$ ``points'' 
as basis vectors.
Working in the computational basis,
we do indeed have $2^n$ such points; 
these are the elements of $\Fnd.$
And so we make the identification
$$
    \Complex[2^n] \cong \Complex[\Fnd].
$$
In other words, we are labeling 
our basis vectors with elements of $\Fnd$
and therefore such notation as
$$
    \bra{u} H \ket{v}
$$
with $u, v \in \Fnd$ makes sense.
We will make further use of this below,
by writing  $\Field$-vector space 
computations inside the Dirac brackets.

Returning to the code $(L,S,T,R)$-decomposition
above,
given the Pauli operator $t\in T$ such that $t = t_X t_Z$ (in $\Pauli_n$)
we get a basis for the irrep $\rho_t$:
$$
    \{ \ket{v R_X + t_X} \ \mbox{such that}\  v \in \Frd \}.
$$
In other words,
the basis of the irrep $\rho_t$ is 
an affine subspace of $\Fnd.$
Each such affine subspace is indexed by an
element of $\Frd$ and
all of these are
translates of each other,
so we make the following identification:
$$
\Complex[\{vR_X+t_X\}_{v\in\Frd}]
\cong \Complex[\Frd].
$$
This will allow us to write the matrix entries
of each block $H_t$ of the Hamiltonian
as $\bra{u}H_t\ket{v}$ for $u,v\in\Frd.$
We make this identification of affine subspaces
not out of laziness but because it will
help us to compare each of
the Hamiltonian blocks $H_t$ below.

\begin{framed}
\noindent
The computational basis identifies
basis vectors of $\Complex[2^n]$
with elements of a finite vector space $\Fnd$:
$$
    \Complex[2^n] \cong \Complex[\Fnd].
$$
The $(L,S,T,R)$-decomposition naturally
decomposes $\Fnd$ into $2^{m_Z+k}$ affine subspaces:
$$
    \{ v R_X + t_X + l_X \}_{v\in\Frd}
$$
for each $t_X \in \Span{T_X}, l_X \in \Span{L_X}.$
Each such affine subspace forms a
symmetry invariant basis
for the irreducible blocks $H_{t_X,t_Z}$ of $H$,
and can be naturally identified with $\Frd:$
$$
    H_{t_X,t_Z} : \Complex[\Frd] \to \Complex[\Frd].
$$
\end{framed}

We now wish to understand the action of the
gauge group on each of its irreps.
Starting with the $t_X,t_Z=0,0$ irrep, each stabilizer has a trivial action. 
In $\Fn$ this
corresponds to the additive action of the zero vector.

\newcommand{\pluseq}{\mathrel{+}=}
States $u\in\Span{R_X}$ can be built from a
vector matrix product
$$
    u = v R_X
$$
with $v\in\Frd.$
Since $R_X R_Z^\top = I$
we can write $v = u R_Z^\top.$
Each $g_X\in G_X$ acts on $u$ to give
\begin{align*}
    u_1 &= (u + g_X) \ \mbox{mod}\ \Span{S_X} \\
        &= (u + g_X) P_X \\
        &= (v R_X + g_X) P_X.
\end{align*}
writing $u_1 = v_1 R_X$ we then have
\begin{align*}
    v_1 &= (v R_X + g_X) P_X R_Z^\top \\
        &= v + g_X R_Z^\top.
\end{align*}
So we have that working in the computational
basis, the action of the $X$ part of the
gauge group in the $t_X,t_Z=0,0$ irrep is to send
$v\in \Frd$ to $v + g_X R_Z^\top.$
In summary, we have the following contributions from the
$G_X$ terms of the Hamiltonian:
$$
    \bbra{v} H_{0,0} \bket{v+g_X  R_Z^\top} 
        \pluseq 1,\ \ \mbox{for}\ g_X\in G_X, v\in \Frd
$$
where we use the $\pluseq$ notation
because there may be other contributions to the
same entry.
These terms will always be off
the diagonal unless $g_X$ is a stabilizer.

The action of the $G_Z$ gauge group
contributes to the diagonal of $H.$
These diagonal terms apply a kind of
``potential energy'' penalty
to the basis states
that depends on the \emph{syndrome} vector:
$$
    \mbox{syndrome}(u) = G_Z u^\top
$$
for $u^\top \in \Fn.$
This is an $\Field$-vector that has one entry for
each row of $G_Z.$
Writing $|G_Z|$ for the number of these rows, and 
using a \emph{weight} function $w$ that just counts
the number of non-zero entries in any $\Field$-vector
we have the following contributions to
the Hamiltonian:
$$
    \bbra{v} H_{0,0} \bket{v} 
        \pluseq |G_Z| - 2w(G_Z R_X^\top v^\top),
$$
for $v\in \Frd.$

Adding up all of the above we
have in summary,
\begin{equation}\label{hamblockgs}
H_{0,0} = \sum_{\substack{v\in\Frd\\g_X\in G_X } }
  \bket{v+g_X  R_Z^\top}\bbra{v} 
  + \sum_{v\in\Frd} \bigl(|G_Z| - 2w(G_Z R_X^\top v^\top)
    \bigr) \ \bket{v}\bbra{v}.
\end{equation}

For any $t_X\in \Span{T_X}$ the Hamiltonian block $H_{t_X,0}$
has entries indexed by basis vectors:
$$
    u = v R_X + t_X
$$
this means that the $G_X$ gauge terms
have the same effect on $H_{t_X,0}$
as $H_{0,0}$ and only the diagonal has changed:
\begin{equation}\label{hamblocktx}
H_{t_X,0} = \sum_{\substack{v\in\Frd\\g_X\in G_X } }
  \bket{v+g_X  R_Z^\top}\bbra{v} 
  + \sum_{v\in\Frd} \bigl(
    |G_Z| - 2w(G_Z R_X^\top v^\top + G_Z t_X^\top)
    \bigr) \ \bket{v}\bbra{v}
\end{equation}
The general form of
each Hamiltonian block is:
\begin{equation}\label{hamblockgen}
H_{t_X,t_Z} = \sum_{\substack{v\in\Frd\\g_X\in G_X } }
    \eta(t_Z g_X^\top)
  \ \bket{v+g_X  R_Z^\top}\bbra{v} 
  + \sum_{v\in\Frd} \bigl(
    |G_Z| - 2w(G_Z R_X^\top v^\top + G_Z t_X^\top)
    \bigr) \ \bket{v}\bbra{v}
\end{equation}
Here we use $\eta$ to send 
$t_Zg_X^\top$ which is an $\Field$ value
to the multiplicative subgroup $\{\pm1\}$
of $\Complex:$
$$
    \eta(0) = 1,\ \eta(1) = -1.
$$
The $\eta(t_Zg_X^\top)$ term
is a kind of parity check that
picks up one phase flip for (some of)
the $X$-type stabilizers found in $g_X.$
This works because $T_Z$ is a left inverse
of $S_X^\top.$
The $t_Z\in\Span{T_Z}$ selects which
$X$-type stabilizers act as $-1$ in this irrep.

In summary, we have the complete representation
theory for $CSS$ gauge code Hamiltonians.

%

\section{Perron-Frobenius theory}\label{Sec5}

In this section we present 
applications of Perron-Frobenius theory to finding bounds on the spectrum of
CSS gauge code Hamiltonians. These are new results.

A \emph{non-negative matrix} is a real matrix
$\Gamma$ with entries $\Gamma_{ij}\ge 0.$
A \emph{positive vector} $\ket{v}$ is a real vector 
with all-positive entries. Similarly, a \emph{non-negative vector}
has all entries real and non-negative.

Given a non-negative matrix 
$\Gamma : \Complex[V] \to \Complex[V]$
we will also view this as a \emph{weighted graph} on a set of vertices $V.$
This associates a directed edge $i\mapsto j$ for every positive
entry $\Gamma_{ij}>0.$

A non-negative matrix 
$\Gamma : \Complex[V] \to \Complex[V]$
is called \emph{irreducible} if the following holds:
for any pair $u, v\in V$ with $u\ne v$ there is
a sequence $\{u_i \}_{i=1,...,n}$ such that $u_1=u$
and $u_n=v$ and
$$
    \bra{u_{i+1}}\Gamma\ket{u_i} > 0\ \ \ \mbox{for}\ \ i=1,...,n-1.
$$
This says that viewed as a weighted graph, $\Gamma$ is ``connected''.
That is, we can find a directed path in $\Gamma$
from any vertex $u$ to any other vertex $v$.
In general, this is a coarser notion than
the (vector space) reducibility used in the representation theory above.
The reason why is that when reducing a matrix we work
with a fixed basis.

The following theorem originally appeared
in the work of Frobenius \cite{Frobenius1912}, who
was building on previous results of Perron \cite{Perron1907}.
See also the recent book \cite{Baez2012}.

\dotheorem{(Perron-Frobenius)}
Let $\Gamma$ be a non-negative irreducible matrix.
Then $\Gamma$ has a positive real eigenvalue
$\lambda$ equal to its spectral radius,
this eigenvalue is non-degenerate and can be associated 
with a positive eigenvector $\ket{v}$.
Furthermore, any other non-negative eigenvector of 
$\Gamma$ is a scalar multiple of $\ket{v}.$
\tombstone

Therefore, when $\Gamma$ is a symmetric non-negative irreducible matrix,
we see that this theorem implies that the top eigenvalue of $\Gamma$ is
non-degenerate and associated with a positive eigenvalue.

Given a matrix $\Gamma$ with $\Gamma_{ij}\ge 0, i\ne j$ we can
find a spectral shift operator which is a constant multiple of the identity $+cI$,
such that $\Gamma+cI$ is a non-negative matrix. 
We refer to such a matrix $\Gamma$ as \emph{stoquastic} \cite{Bravyi2008,Bravyi2015}.
When $\Gamma$ is also irreducible we can therefore apply the Perron-Frobenius
theorem with the proviso that the eigenvalues are shifted by some constant.

For a linear operator $A$ we will use notation 
$\lambda_1(A)$, $\lambda_2(A),$ etc., to denote the (real) eigenvalues of $A$,
ordered as:
$$
    \lambda_1(A) \ge \lambda_2(A) \ge ...
$$
Similarly, the notation $v_1(A), v_2(A),$ etc., refers to the associated eigenvectors.



Given a CSS gauge code Hamiltonian $H,$ 
let the matrix for $H$ in the computational basis be $\Gamma.$
We see that the diagonal entries of $\Gamma$
come from the $Z$-type operators and the off-diagonal entries 
come from $X$-type operators.
Therefore, the off-diagonal entries of $\Gamma$ are non-negative 
and so this matrix is stoquastic.
In the language of graph theory,
$\Gamma$ has vertices the $2^n$ computational
basis elements, and edges:
$$
    \ket{v} \mapsto g_X \ket{v}, \smbox{for all} v\in\Fnd, g_X\in G_X.
$$
These are undirected edges because $g_X^2 = I.$
We also add weighted loops corresponding to the $Z$-type gauge operators:
$$
    \ket{v} \mapsto \sum_{g_Z\in G_Z} g_Z \ket{v}, \smbox{for all} v\in\Fnd.
$$

We can think of the $Z$-type operators as potential energy,
and the $X$-type operators as kinetic energy.
This is consistent with the classical use of the (real) symplectic group,
where dynamical variables come in position and momentum pairs.

\dolemma{\numitem{lemma21}}
Given a CSS gauge code Hamiltonian $H$, writing
the matrix for $H$ in the computational basis as $\Gamma,$
this matrix decomposes into irreducible
stoquastic matrices $\Gamma_{t_X}$ 
indexed by $t_X\in\Span{T_X}$
with multiplicity $2^k$:
$$
    \Gamma = \bigoplus_{
    \substack{t_X\in\Span{T_X}\\l_X\in\Span{L_X}}}
        \Gamma_{t_X}.
$$

\doproof
Once again, we identify the computational basis with
the elements of $\Field_n.$
As a graph, $\Gamma$ has vertices 
$\ket{v} $ for $v\in\Field_n$, and edges 
$ \ket{v} \mapsto g \ket{v}$ for  $v\in\Fnd, g\in G_X.$
This means that paths starting from $\ket{v}$ are given by
words $g_i...g_1$ built from the elements of the group $G_X:$
$$
    \ket{v} \mapsto g_1 \ket{v} \mapsto ... \mapsto g_i...g_1 \ket{v}.
$$
Switching to the additive $\Field_n$ notation, we
therefore have that 
vertices $\ket{v}$ and $\ket{u}$ are in the
same component of $\Gamma$ iff $\ket{u} = \ket{v+g_X}$ for some $g_X\in\Span{G_X}.$
The vertices
$$\Bigl\{ \bket{t_X + l_X} \Bigr\}_{t_X\in\Span{T_X},\ l_X\in\Span{L_X}}$$
each live in one component of $\Gamma$,
and each component of $\Gamma$ contains one of these vertices.
This shows that these components are in bijective correspondence with
pairs $(t_X, l_X)$ where $t_X\in\Span{T_X},\ l_X\in\Span{L_X}$.
We write each such component as $\Gamma_{t_X,l_X}$.
The vertices contained in the $\Gamma_{t_X,l_X}$ component
are given by a coset of $G_X$ in $\Fnd:$
$$
    \Bigl\{ \bket{v S_X + u R_X + t_X + l_X} \Bigr\}_{v\in \Field_{m_X}, u\in\Frd}.
$$
When comparing these different components of $\Gamma$, we will
identify these cosets, and so consider $\Gamma_{t_X,l_X}$ as an
operator:
$$
    \Gamma_{t_X,l_X} : \Complex[\Field_{m_X}\oplus\Field_r] \to \Complex[\Field_{m_X}\oplus\Field_r].
$$
Because each of these components is independant of $l_X$ we write it as $\Gamma_{t_X}.$
\tombstone




\dolemma{\numitem{lemma22}}
Given a CSS gauge code Hamiltonian $H$, 
and matrix $\Gamma$ in the computational basis 
with irreducible
components $\Gamma_{t_X}$,
with $t_X\in\Span{T_X}$,
the first eigenvalue 
of each $\Gamma_{t_X}$
is non-degenerate
and is associated with an eigenvector 
with positive entries.

\doproof
We already established that $\Gamma$ and therefore
$\Gamma_{t_X}$ is stoquastic.
By the previous lemma $\Gamma_{t_X}$ is irreducible.
We apply the Perron-Frobenius theorem and the result follows.
\tombstone

Any stabilizer that acts as $-1$ in a given block of
the Hamiltonian we will call a \emph{frustrated stabilizer}
(with respect to the given block.)
Similarly, a \emph{satisfied stabilizer} acts as $+1$ in a given
block of the Hamiltonian.
We will call a
vector \emph{stabilized} if it is a $+1$ eigenvector 
of every stabilizer in the given gauge group.

\dolemma{\numitem{lemma23}}
Given a CSS gauge code Hamiltonian $H$,
every groundstate of $H$ is stabilized.

\doproof
The set of top eigenvectors 
$$
    \bigl\{ v_1(\Gamma_{t_X}) \bigr\}_{t_X\in\Span{T_X}}
$$
must contain a basis for the groundspace of $H.$
Let such a basis vector be $\ket{v} = v_1(\Gamma_{t_X})$ for some $t_X.$
This vector is positive by Lemma \ref{lemma22}.
Any stabilizer commutes with
every element of the gauge group and so
must have $\ket{v}$ as an eigenvector.
Because any
$X$-type operator acts by permuting the computational basis elements 
and $\ket{v}$ is positive it must be stabilized by the
$X$-type stabilizers.
Therefore, we can find a basis for the groundspace of
$H$ consisting of vectors that are fixed by $X$-type
stabilizers.

We repeat this argument for $H$ switching to the
$\ket{\pm}$ basis. This will give a basis for the groundspace
consisting of vectors that are fixed by the $Z$-type
stabilizers.
Therefore the entire groundspace must be stabilized.
\tombstone

\doproposition{\numitem{prop24}}
For any CSS gauge code Hamiltonian $\Ham$
we have:
$$\lambda_1(\Ham) = \lambda_1(\Ham_{0,0})$$
and for $t_X\in\Span{T_X}, t_Z\in\Span{T_Z}$
with $t_X\ne 0$ or $t_Z\ne 0$,
$$
\lambda_1(\Ham) > \lambda_1(H_{t_X,t_Z}).
$$

\doproof
The vectors in the space that 
$H_{t_X,t_Z}$ operates on 
are stabilized iff $t_X=0$ and $t_Z=0$,
and so the result follows as a 
consequence of the previous lemma.
\tombstone

The following lemma is used in section 2.8 below.

\dolemma{\numitem{lemma25}}
Given a 
CSS gauge code Hamiltonian $\Ham$ and
$t_X\in \Span{T_X},$
the matrices for $\Ham_{t_X,0}$ 
in the symmetry invariant basis 
are irreducible stoquastic.

\doproof
From equation \Eref{hamblocktx} we see that 
the nonzero, nondiagonal entries of the matrices
$\Ham_{t_X,0}$ are $+1,$ and so this matrix is stoquastic.
This matrix is irreducible because as an operator it is
irreducible.
\tombstone

So far, the Perron-Frobenius theory has been fruitful,
telling us that the first eigenvalue of $H$ resides in
the spectrum of the block $H_{0,0}$.
The next goal is to search for
the second eigenvalue of $H$.

\dolemma{\numitem{lemma26}}
Let $H$ be a CSS gauge code Hamiltonian,
with matrix $\Gamma$ in the computational basis 
and irreducible components $\Gamma_{t_X}$.
We have
$$
    \Gamma_{t_X} = \bigoplus_{t_Z\in\Span{T_Z}}
        H_{t_X,t_Z}.
$$

\doproof
Irreducibility of matrices is a coarser property than irreducibility of 
representations. 
The result follows by considering eigenspaces of stabilizers.
\tombstone

\doproposition{\numitem{prop27}}
For a gauge code Hamiltonian $H$,
\begin{align*}
\lambda_1(\Ham_{t_X,0}) &> 
    \lambda_1(\Ham_{t_X,t_Z}) \ \smbox{and}\\
\lambda_1(\Ham_{0,t_Z}) &> 
    \lambda_1(\Ham_{t_X,t_Z}),
\end{align*}
where $t_X\ne 0$ and $t_Z\ne 0.$

\doproof
Let $H$ be a CSS gauge code Hamiltonian.
Let $\Gamma$ be the matrix for $H$ in the computational basis,
with $\Gamma_{t_X}$ the irreducible matrix components.
From Lemma \ref{lemma26},
$$
    \Gamma_{t_X} = \bigoplus_{t_Z\in\Span{T_Z}}
        H_{t_X,t_Z}.
$$
And, by Lemma \ref{lemma22} the first eigenvector of $\Gamma_{t_X}$ is 
non-degerate and positive.
This will be an eigenvector of each stabilizer, and in particular
must be fixed by the $X$-type stabilizers.
Therefore, this eigenvalue is in the spectrum of $H_{t_X,0}$ and
no other $H_{t_X,t_Z}$ with $t_Z\ne 0.$
This shows 
$\lambda_1(\Ham_{t_X,0}) > \lambda_1(\Ham_{t_X,t_Z}).$
We switch to the $\ket{\pm}$ basis and repeat this argument
to obtain the second inequality.
\tombstone

To find the spectral gap of a weakly self-dual gauge code Hamiltonian $H$,
we need only examine the top two eigenvalues of $H_{0,0}$ and 
the top eigenvalue of $H_{t_X,0}$ for each $t_X\in T_X.$ 
We summarize this in the theorem:

\dotheorem{\numitem{thm28}}
For a weakly self-dual gauge code Hamiltonian $H$,
the spectral gap is given by:
$$
    \min \bigl\{ \lambda_1(H_{0,0}) - \lambda_2(H_{0,0}),
        \min_{t_X\in\Span{T_X}} 
         \{ \lambda_1(H_{0,0}) - \lambda_2(H_{t_X,0}) \} \bigr\}
$$
\doproof
Combine Proposition \ref{prop24} with Proposition \ref{prop27}.
By weak self-duality, the spectrum of $H_{t_X,t_Z}$ equals
the spectrum of $H_{t'_X,t'_Z}$ for some 
$t'_X\in\Span{T_X}, t'_Z\in\Span{T_Z}.$
\tombstone

%

\section{Models}\label{Sec4}

A CSS gauge code is \emph{self-dual} when the $X$-type and $Z$-type
gauge generators are equal: $$G_X = G_Z.$$
A CSS gauge code is \emph{weakly self-dual}
when there is a permutation $P$ on the set of $n$ qubits
that induces equality of the gauge generators:
$$
    G_X P = G_Z,
$$
where we write $P$ as an $n\times n$ permutation matrix.

\subsection{Transverse field Ising model}\label{Sec42}

For the one dimensional transverse field
Ising model, we have 
$$
    G_0 = \{ X_i, Z_i Z_{i+1}\ \ \mbox{for}\ \ i=1,...,n \},
$$
with periodic boundary conditions.
This model has one extensive 
$X$-type stabilizer and no logical operators.
This model is also exactly solvable, with gap going to zero
as the system size grows \cite{Pfeuty1970}.

\subsection{$XY$-model}\label{Sec41}

The $XY$-model \cite{Pfeuty1970}
lives on a one dimensional chain of $n$ qubits.
We write the gauge group generators as
$$
    G_0 = \{ X_i X_{i+1}, Z_i Z_{i+1}\ \ \mbox{for}\ \ i=1,...,n \}
$$
with periodic boundary conditions.
This is a self-dual CSS gauge code.
We focus on the $n$ even $XY$-model,
which is exactly solvable,
with gap going to zero as the system size grows \cite{Lieb1961}.
This model has no logical operators, one 
$X$-type stabilizer and one $Z$-type stabilizer,
both extensive.
Normally the gauge operators are written as 
$\{ X_i X_{i+1}, Y_i Y_{i+1} \ \mbox{for}\ i=1,...,n \}$
but note that there is an automorphism of the Pauli group
that sends $G_0$ to these operators.

We also consider a two dimensional version of this
model, which we call the \emph{2D $XY$-plaquette model}.
Qubits live on a 2D periodic lattice of $l\times l$ sites.
This is also a self-dual CSS gauge code, with 
gauge group generators 
$$
    G_0 = \{
        X_{ij}X_{i+1,j}X_{i,j+1}X_{i+1,j+1},\ 
        Z_{ij}Z_{i+1,j}Z_{i,j+1}Z_{i+1,j+1}
        \ \ \mbox{for}\ \ i,j=1,...,l \}.
$$
For $l$ even, we have $2l-3$ extensive stabilizer generators of $X$ type:
$$
    \prod_{i=1}^l X_{ij} X_{i,j+1} \ \ \mbox{for}\ \ j=1,...,l-1,\ \ 
    \prod_{j=1}^l X_{ij} X_{i+1,j} \ \ \mbox{for}\ \ j=1,...,l-2,
$$
and similarly for the $Z$-type stabilizer generators.
This model has two logical qubits with logical operators 
generated by
$$
    L_0 = \Big\{ 
            \prod_{i=1}^l X_{i1}, \ 
            \prod_{i=1}^l Z_{i1}, \ 
            \prod_{j=1}^l X_{1j},\ 
            \prod_{j=1}^l Z_{1j}
        \Big\}.
$$

\subsection{Compass model}\label{Sec43}

Here we consider the compass model \cite{Bacon2006}.
For the 2D model,
qubits live on a periodic square lattice of $l\times l$ sites.
The generators of the gauge group are
$$
    G_0 = \big\{ X_{ij}X_{i,j+1},\ Z_{ij}Z_{i+1,j}\ \
        \mbox{for}\ \ i,j = 1,...,l\big\}.
$$
The generators for the stabilizers are
extensive and given by
$$
    \Stab_0 = \Big\{ \prod_{i=1}^l X_{ij}X_{i,j+1},\ \prod_{i=1}^l Z_{ji}Z_{j+1,i}
        \ \ \mbox{for}\ \ j=1,...,l-1\Big\}.
$$
The logical operators are generated by 
$L_0 = \big\{ \prod_i X_{i1}, \prod_j Z_{1j} \big\}.$
This model is weakly self-dual when we transpose
the square lattice of $l\times l$ qubits.

The 3D compass model has qubits on a periodic lattice of $l\times l\times l$
sites.
The generators of the gauge group and stabilizer group are
\begin{align*}
    G_0 &= \big\{ 
        X_{ijk}X_{i+1,j,k},\ 
        X_{ijk}X_{i,j+1,k},\ 
        Z_{ijk}Z_{i,j+1,k},\ 
        Z_{ijk}Z_{i,j,k+1}\ 
        \ \mbox{for}\ \ i, j, k = 1,...,l\big\}\\
    \Stab_0 &= \Big\{ 
        \prod_{j,k=1}^l X_{ijk}X_{i+1,j,k},\ 
        \prod_{j,k=1}^l Z_{jki}Z_{j,k,i+1}\ \ 
            \mbox{for}\ \ i=1,...,l-1\Big\}.
\end{align*}

The 3D compass model is also weakly self-dual,
and has extensive stabilizer generators.

\subsection{Gauge color code model}\label{Sec44}

The three dimensional gauge color code \cite{Bombin2015,Bombin2015single,Kubica2015}
is a self-dual CSS gauge code. 
It is based on the following geometric construction known
as a \emph{colex} \cite{Bombin2007exact}.
We begin with a tetrahedron and subdivide it into finitely many
convex 3-dimensional polytopes, or \emph{bodies}.
Each body has a boundary consisting of 0-dimensional cells
which we call \emph{vertices}, 1-dimensional cells called \emph{edges}
and 2-dimensional cells called \emph{faces}.
By a \emph{cell} we mean any of these 0,1,2 or 3-dimensional convex polytopes.
Any two bodies in this tetrahedral subdivision will
have either empty intersection or otherwise intersect
on a common vertex, edge or face.
When the intersection is on a face these two bodies
are called \emph{adjacent}.
Two vertices in the same edge will also be called adjacent.
Each body is colored by one of four \emph{colors},
either taken to be red, green, blue, yellow or 
otherwise an element of the set $\{1, 2, 3, 4\}.$
The four exterior triangular faces of the bounding tetrahedron are
called \emph{regions,}
and the intersection of three regions is called
a \emph{corner.}
A cell not contained within any region is called an interior cell.

This colored cellulation is required to have the following further properties:
\begin{enumerate}
\item Adjacent bodies have different colors.
\item Each region has a unique color 
such that no bodies intersecting that region has that color.
\item All vertices are adjacent to four other vertices,
except for the corner vertices which are adjacent to three other vertices.
\end{enumerate}

Here we show some instances of this construction,
along with the colors of the unobscured bodies.
Each instance is labeled by the number of vertices $n$.
\begin{center}
\includegraphics{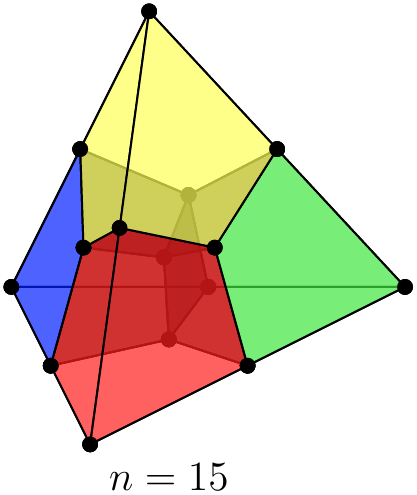}\ \ \ \ \ \ \ \ \ \ \ 
\includegraphics{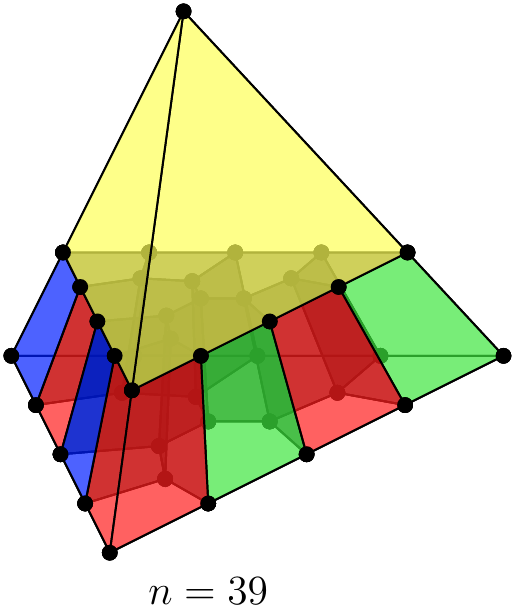}
\end{center}

\begin{center}
\includegraphics{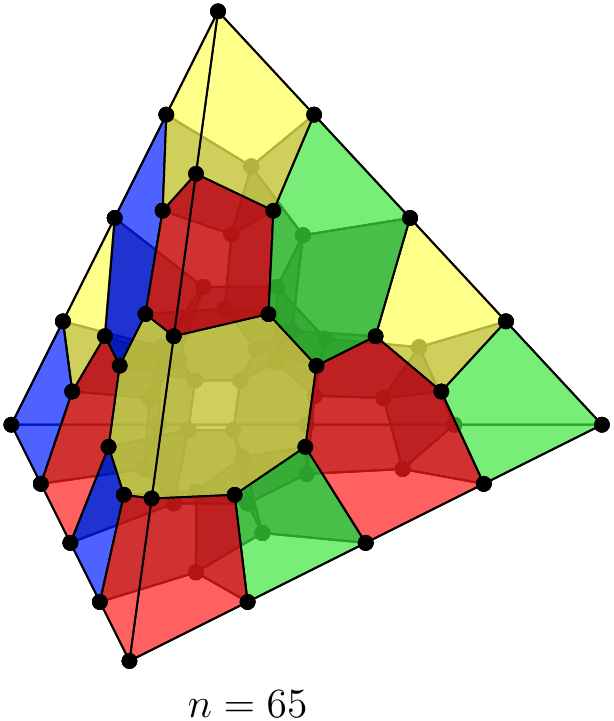}\ \ \ \ \ \ \ \ \ \   \includegraphics{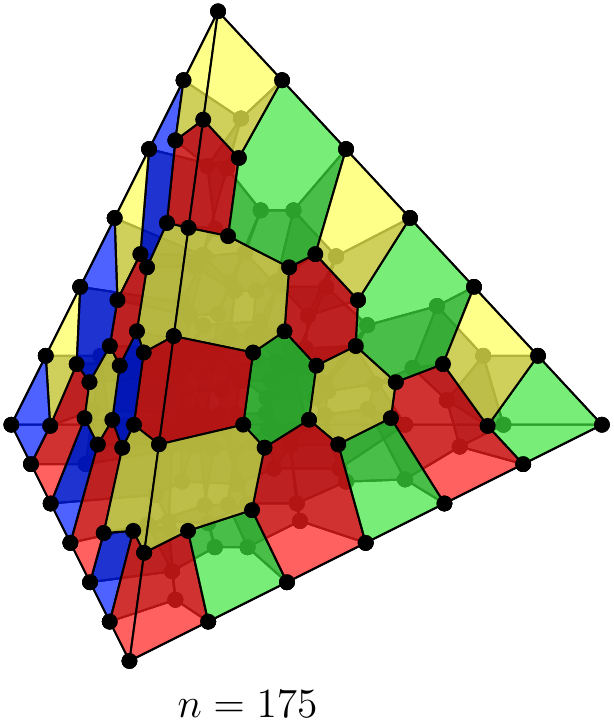}
\end{center}

We note the following consequences of the above conditions.
Every face supports an even number of vertices.
Thinking of each region as corresponding to a missing body,
each edge is then contained in the boundary of three bodies.
This means we can associate a unique color to each edge,
which is also the color of two bodies intersecting a vertex of the edge.
That is, each edge joins two bodies of the same color.
Each face bounds two bodies, and so we color each face with the two colors of these bodies.

Using this cellulation we now construct the gauge code.
Qubits are associated with the $n$ vertices.
We associate operators to other cells, or union of cells,
by using the contained vertices as support.
Because this is a self-dual code, the same definitions apply for
both the $X$-type and $Z$-type operators.
The $X/Z-$type gauge group is generated by
$X/Z-$type operators supported on each face.
The $X/Z-$type stabilizer group is generated by
$X/Z-$type operators supported on each body.
Therefore, the stabilizer generators are non-extensive
(have bounded size).
There is one $X/Z-$type logical operator generator and these
are given by $X/Z-$type operators supported on any region.

\subsection{Commuting ideals of gauge codes}\label{Sec45}

When the gauge group generators $G_0$ can be
partitioned into mutually commuting parts, the
corresponding Hamiltonian can be written as a sum
of mutually commuting operators.
For example,
for $n$ even, the $XY$-model gauge generators can
be partitioned into two mutually commuting parts:
\begin{align*}
G_0 = &\{X_i X_{i+1}, Z_i Z_{i+1}\  \smbox{for} i=1,...,n. \} \\
    = &\{X_i X_{i+1}, Z_{i+1} Z_{i+2}\ \smbox{for}  i=1,3,..,n-1. \}\\
    & \cup \{X_i X_{i+1}, Z_{i+1} Z_{i+2}\ \smbox{for}  i=2,4,..,n. \}.
\end{align*}
In general we call these parts \emph{commuting ideals}.\footnote{
These parts generate Lie algebra ideals when we consider 
the usual commutator $[\cdot,\cdot]$ as Lie algebra bracket.}
Let $a\in A$ index these parts. 
For a partition of the gauge group into mutually commuting parts $G_0^{a}$,
we write the Hamiltonian $H$ as a sum of mutually commuting
operators $H^a = \sum_{g\in G_0^{a}} g:$
$$
    G_0 = \bigcup_{a\in A} G_0^{a},\ \  
    H^a = \sum_{g\in G_0^{a}} g,\ \  
    H = \sum_{a\in A} H^a.
$$
%
Applying the Decomposition Theorem to each of the
$H^a$ we obtain the matrices $L^a, S^a, T^a, R^a$ with $X,Z$ subscripts.
It follows that
each Hamiltonian block $ H_{t_X,t_Z} $
can be written as a sum of mutually commuting operators:
$$
    H_{t_X,t_Z} = \sum_a H^a_{t_X S_Z^{a\top} T_X^a, t_Z S_X^{a\top} T_Z^a}.
$$

We next examine the commuting ideal structure of the 3D gauge color code.
Two face operators, of $X$-type and $Z$-type,
will anti-commute precisely when they intersect on a single vertex.
And this only happens when such faces have disjoint coloring.
Here we show an example of this in the $n=39$ model:
\begin{center}
\includegraphics{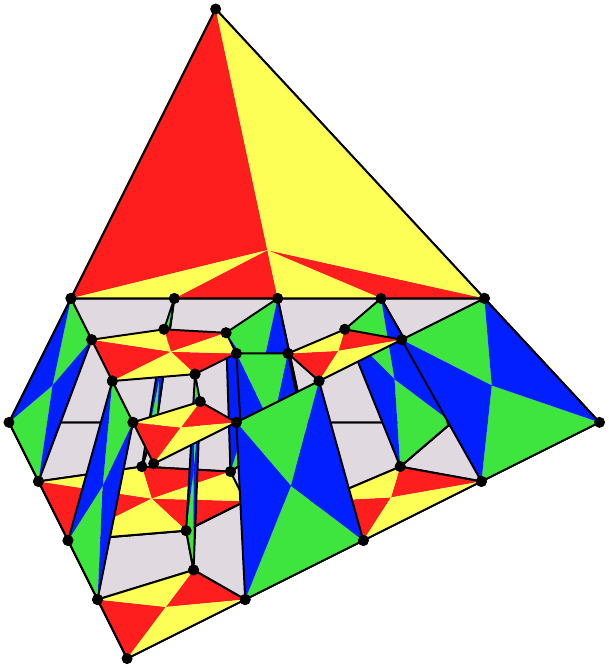}
\end{center}
There are three of these arrangements,
each corresponding to the three ways of
partitioning the set of colors into two sets of two.
It follows that $G_0$
can be partitioned into six commuting ideals.

In a similar vein,
for the 2D $XY$-plaquette model (with $l$ even)
we find that the gauge group partitions
into four commuting ideals.
Each ideal corresponds to a choice of 
a checkerboard coloring of the $l\times l$ square lattice,
and an assignment of either $X$-type or $Z$-type
gauge group operators to the black squares.
The white squares then support the other type of operator.

The results in this section are crucial for obtaining the exact
diagonalization numerical results below.
By partitioning the gauge group in this way, 
we obtain a partition on the reduced gauge group. 
This gives an exponential reduction
in the numerical workload.


%


\section{Numerical results}\label{Sec6}

Here we show tables for the first and
second eigenvalues of the 2D and 3D compass models,
as well as the 2D $XY$-plaquette model and gauge color code model.
These results are obtained using exact diagonalization methods.
For each instance we indicate the groundspace eigenvalue
$\lambda_1$ which is obtained from $H_{0,0}.$
Then we list the second eigenvalue of $H_{0,0}$ as
well as the first eigenvalue of $H_{t_X,0}$ for $t_X\ne 0.$
The weight of the corresponding frustrated stabilizer is $w(s_Z).$
The eigenvalue closest to $\lambda_1(H_{0,0})$ is marked
with a tick, along with the value of the gap, $\lambda_1-\lambda_2.$
We only show the results for a single frustrated
stabilizer generator,
as it was confirmed numerically that 
adding further frustrated stabilizers never 
produces a better candidate for $\lambda_2.$
This involved performing exact diagonalization on 
the top eigenvalues of every $H_{t_X,0}$ block in the Hamiltonian.
Also, we only show non-isomorphic stabilizer generators,
under the lattice symmetry of the model.
We use the iterative solvers in the software library 
{\tt SLEPc} \cite{Hernandez2005} to find these eigenvalues.

\begin{samepage}
\underline{2D compass model}
\begin{center}
\begin{tabular}{ c|c|c|c|l|c } 
$n$ &  $t_X$    & $w(s_Z)$ & $\lambda_1$ & $\ \ \ \ \lambda_2$ ? & gap \\
\hline
\hline
16  &   0        &   &  19.012903&    16.335705          &            \\
&            & 8 &              &  18.369300    \checkmark & 0.643603 \\
\hline
25  &   0        &   & 29.076200 & 27.597280        &            \\
&            & 10 &              & 28.624004 \checkmark &  0.452196 \\
\hline
36  &   0        &   & 41.410454 & 40.585673        &            \\
&            & 12 &              & 41.094532 \checkmark &  0.315922 \\
\end{tabular}
\end{center}
\end{samepage}

Such numerics for the 2D compass model have been previously found 
using similar methods \cite{Brzezicki2013}. 
The following numerical results are new:

\begin{samepage}
\underline{3D compass model}
\begin{center}
\begin{tabular}{ c|c|c|c|l|c } 
$n$ &  $t_X$    & $w(s_Z)$ & $\lambda_1$ & $\ \ \ \ \lambda_2$ ? & gap \\
\hline
\hline
27  &   0        &   & 60.295471  &    58.382445          &            \\
&            & 18 &              &  59.757677   \checkmark & 0.53779 \\
\end{tabular}
\end{center}
\end{samepage}

\begin{samepage}
\underline{2D $XY$-plaquette model}
\begin{center}
\begin{tabular}{ c|c|c|c|l|c } 
$n$ &  $t_X$    & $w(s_Z)$ & $\lambda_1$ & $\ \ \ \ \lambda_2$ ? & gap \\
\hline
\hline
16  &   0        &   & 22.627417 & 11.313708      &            \\
&            & 8 &              & 19.313708   \checkmark &  3.31371  \\
\hline
36  &   0        &   & 44.8444102 & 39.633308        &            \\
&            & 12 &              & 42.914196   \checkmark &  1.93021  \\
\hline
64  &   0        &   &  76.051613  &    73.374415    &            \\
&            & 16 &              &    74.764406  \checkmark & 1.28720  \\
\hline
100  &   0        &   & 116.304800 & 114.825880        &            \\
&            & 20 &              & 115.400408   \checkmark &  0.90439  \\
\hline
144  &   0        &   & 165.641816   & 164.817035  &            \\
&            & 24 &              & 165.009972   \checkmark &  0.63184   \\
\end{tabular}
\end{center}
\end{samepage}

\begin{samepage}
\underline{3D gauge color code}
\begin{center}
\begin{tabular}{ c|c|c|c|l|c } 
$n$ &  $t_X$    & $w(s_Z)$ & $\lambda_1$ & $\ \ \ \ \lambda_2$ ?  & gap \\
\hline
\hline
15  & 0         & &  25.455844  & 16.970563    &  \\
                 &   & 8 &              & 22.214755 \checkmark & 3.241089           \\
\hline
65  & 0         &    &  104.076026  & 99.014097     &    \\
                 &           & 8  &              &  100.429340   &            \\
                 &           & 12 &              &  100.585413   &            \\
                 &           & 12 &              &  101.602340   &            \\
                 &           & 18 &              &  102.382483  \checkmark  &  1.693543 \\
\hline
175 & 0         &  &  267.197576  & 264.250644    & \\
 & & 8  & & 263.171190  &    \\
 & & 8  & & 263.324858  &    \\
 & & 8  & & 263.340832  &    \\
 & & 12 & &  264.269635  &    \\
 & & 12 & &  264.617135  &    \\
 & & 12 & &  264.745548  &    \\
 & & 18 & &  264.843629  &    \\
 & & 18 & &  265.413935  &    \\
 & & 18 & &  265.754772  &    \\
 & & 24 & &  266.148188  \checkmark &  1.04939  \\
\end{tabular}
\end{center}
\end{samepage}

\begin{figure*}
\begin{center}
\includegraphics[width=0.8\columnwidth]{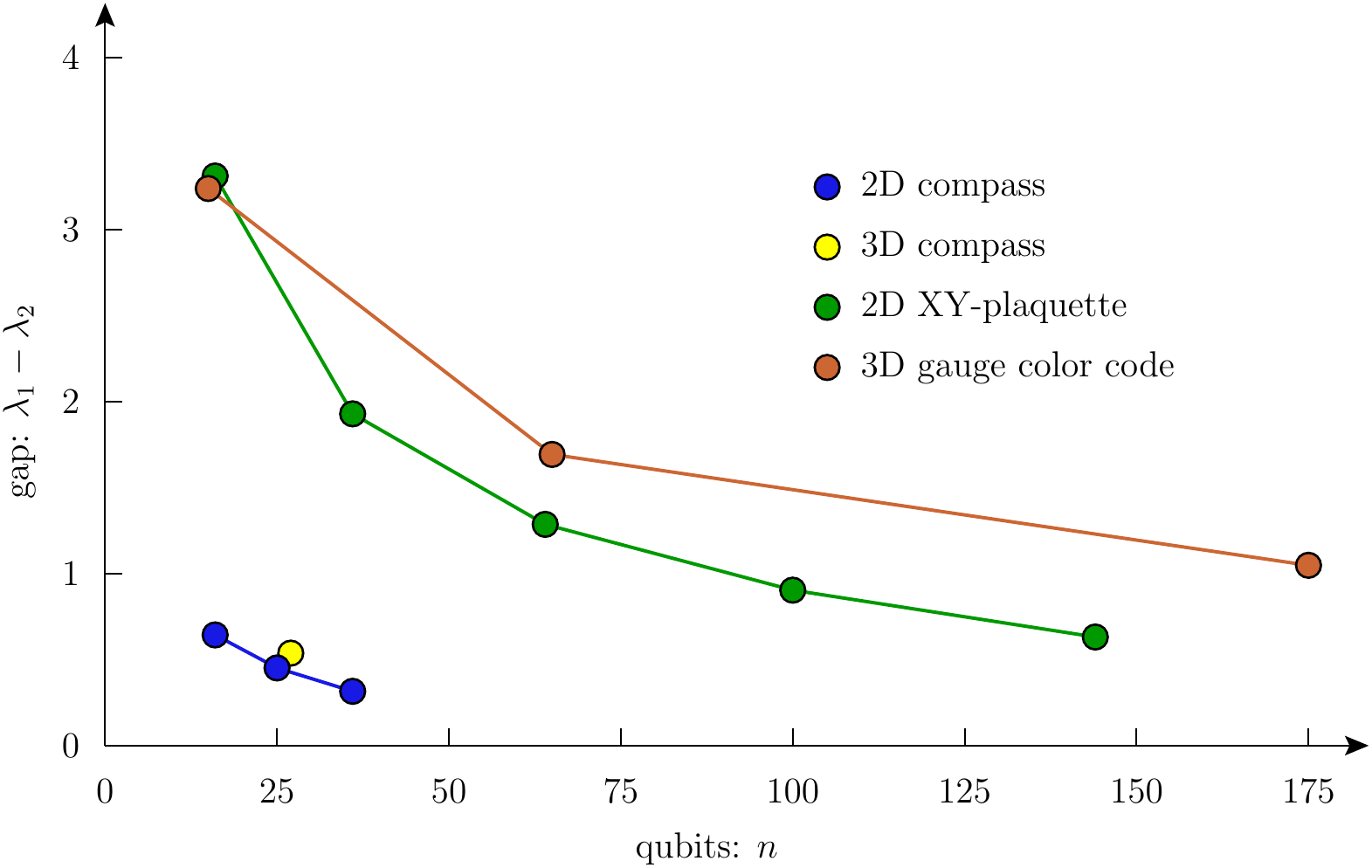}
\caption{%
The spectral gap of four different gauge code Hamiltonians, versus the number
of qubits $n$. The gap is defined as the difference between
the ground eigenvalue and the first excited eigenvalue.
These results are obtained by exact diagonalization.
}
\label{PicGap}
\end{center}
\end{figure*}

\begin{figure*}
\begin{center}
\includegraphics[width=0.8\columnwidth]{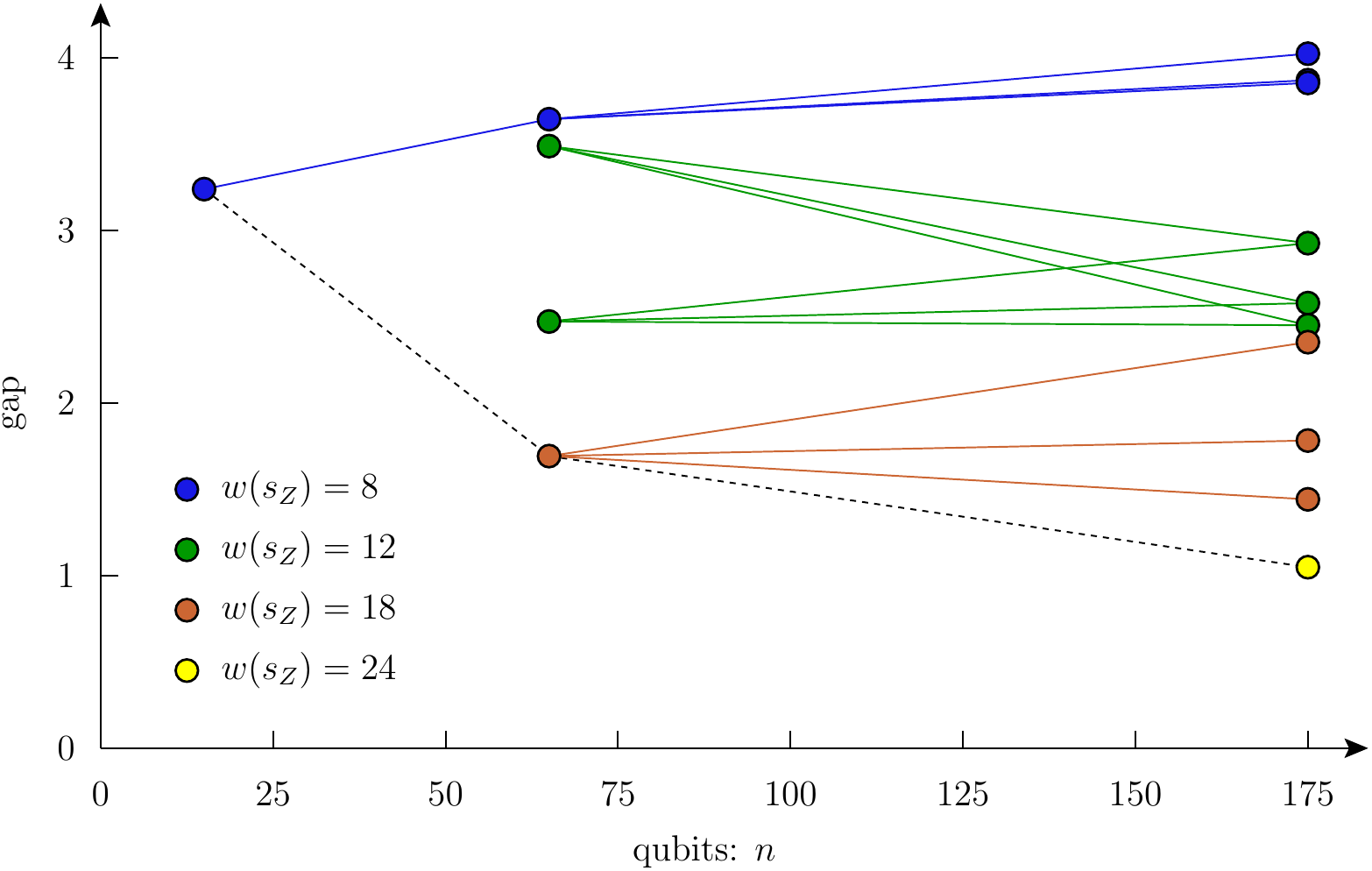}
\caption{%
For the 3D gauge color code model
we show the spectral gap
of each Hamiltonian block $H_{t_X,0}$
with $n=15,65,175.$
This gap is defined as $\lambda_1(H) - \lambda_1(H_{t_X,0}).$
Each point is colored according to the weight of the
frustrated stabilizer generator.
}
\label{PicGapStabs}
\end{center}
\end{figure*}

These numerics are shown in Figures \ref{PicGap} and
\ref{PicGapStabs}.
The gap of the 3D gauge color code is more
robust than the other models, see Figure \ref{PicGap}.
It does decrease with $n$, but note also that the
stabilizers in the code are also growing, up to weight 24.
To emphasize this point we show in Figure \ref{PicGapStabs}
the ground eigenvalues of all of the blocks $H_{t_X,0}.$
For larger codes in this family the stabilizers do not get
bigger than weight 24.
It is not clear to what extent these results are
representative of larger code sizes, but we
can already see from Figure \ref{PicGapStabs} 
evidence that 
the weight of the frustrated stabilizer generator plays a more
important role than the size of the code itself.
In these examples, the gap always corresponds to frustrating
a stabilizer ($t_X\ne 0$)
and in particular the stabilizer generator with largest weight.%
\footnote{There is some subtlety about what we mean by 
stabilizer generator, but hopefully this is clear for the
codes we study here.}
This is a crucial connection to make because the
stabilizers of the compass model 
(as well as the $XY$ and Ising models)
grow with the linear
size of the model
while those of the gauge color code 
do not grow beyond a constant bound.
This would suggest that if this is the mechanism for
gapless behaviour that the gauge color code model may
be gapped.

These numerical results reach the limit of presently available 
computational resources.
To find eigenvalues for each Hamiltonian block $H_{t_X,0}$
we need to operate on wavefunctions with $2^r$ real coefficients.
The iterative solvers in {\tt SLEPc} need to store at least two,
but ideally more, of these wavefunctions.
For $r=32$ this is about 32 Gigabytes for one wavefunction (using double
precision coefficients) and so this value for $r$ is roughly the
upper limit on these numerical techniques.
Without decomposing the gauge color code into six disjoint ideals
it would be impossible to obtain the results for the $n=175$ code,
as this code has $r=94$.

\section{Cheeger cuts}\label{Sec7}

In this final section of this paper 
we give some heuristic
arguments for why the size of the stabilizers is
related to the gap of the Hamiltonian. 

The Perron-Frobenius structure theory places
strong constraints on the first and second
eigenvectors of $\Gamma_{t_X}:$
the first eigenvector has all positive entries,
and therefore all vectors orthogonal to the first
eigenvector will have both positive and negative entries.
In general, the set of edges of $\Gamma_{t_X}$ where
such a vector changes sign we call a Cheeger cut 
\cite{Cheeger1970,Chung1997}.
(We ignore the possibility that this vector
may have zero entries.)
The Cheeger cut associated to the second eigenvector
is particularly important, and we next show an
example of how this cut relates to the gap.

\subsection{The double well model is gapless}\label{Sec71}

We consider a linear graph Hamiltonian
with a ``double-well'' potential.
This does not correspond to any gauge code Hamiltonian.
The state space will be $d$ dimensional with
basis vectors numbered $\ket{1},...,\ket{d}.$
We take
$ \Ham = A + U $
with
$$
A_{ij} = \left\{ \begin{array}{ll}
     1 &\mbox{if}\  |i-j|=1,  \\
     0 &\mbox{otherwise}\end{array}\right.
\ \ \mbox{and}\ \ 
U_{ij} =  \left\{ \begin{array}{ll}
     2 &\mbox{if}\  i=j=1 \ \ \mbox{or}\ \  i=j=n, \\
     0 &\mbox{otherwise.}\end{array}\right.
$$
$A$ here is a kind of transition matrix,
and $U$ is a diagonal potential energy term.

For $d\gg 1$, the largest
eigenvalue is $\lambda_1 \cong \frac{5}{2}$.
The corresponding eigenvector $\ket{v_1}$
has all positive entries that
decay exponentially away from the well sites
at $\ket{1}$ and $\ket{d}:$
$$
    \braket{i}{v_1} 
    \cong 2^{i-1} \braket{1}{v_1}
    \ \ \mbox{for}\ \ i\ll \frac{d}{2}.
$$
For the second eigenvalue, $\lambda_2$
we also have  $\lambda_2 \cong \frac{5}{2}$
and indeed, as $d$ grows
the gap $\lambda_1 - \lambda_2 \rightarrow 0$
and so this model is gapless.

Here we depict the wavefunctions for
the first two eigenvectors for a system with $d=12:$
\begin{center}
\includegraphics[]{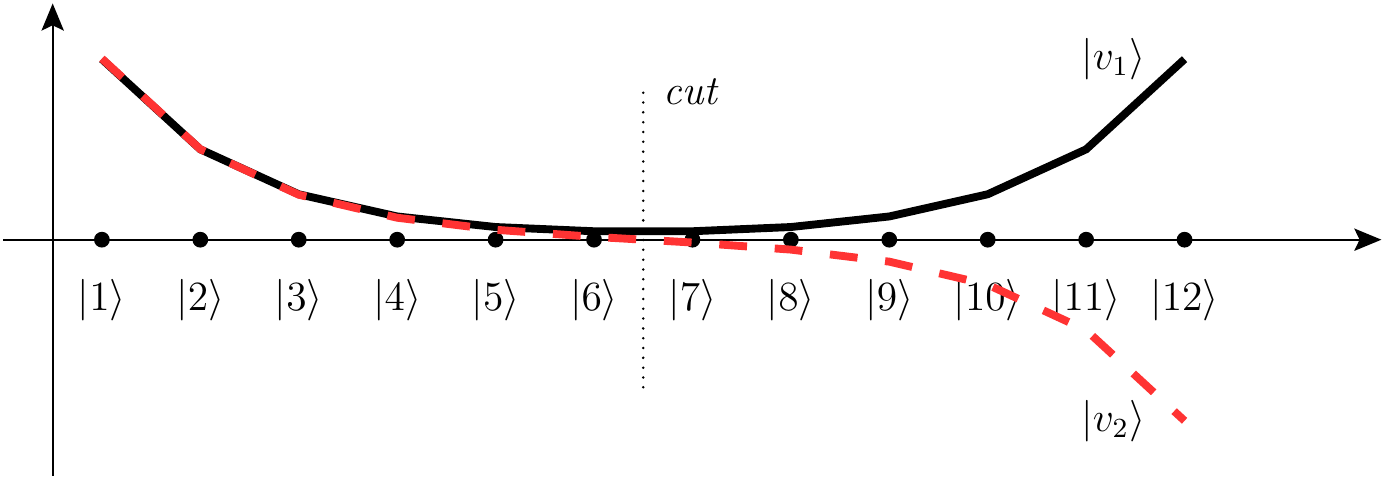}
\end{center}
The simplest way to show this model
is gapless is using a variational
argument.
Any another vector $\ket{u}$
that is orthogonal to the groundspace
vector will have $\bra{u}H\ket{u} \le \lambda_2.$
To construct a candidate for $\ket{u}$
partition the
basis vectors into two parts:
$$
    \Gamma = \Gamma_A \cup \Gamma_B
$$
and write $\ket{v_1} = 
\ket{v_A}\oplus\ket{v_B}$
as well as Hamiltonian with this
decomposition as
$$
H = 
\left(\begin{array}{ll}
H_{AA} & H_{AB} \\
H_{BA} & H_{BB}
\end{array}\right).
$$
Now let
$$
    \ket{u} = \ket{v_A} \oplus -\ket{v_B}
$$
And then
\begin{align*}
    \lambda_2 \ge \bra{u}H\ket{u} &= 
\bra{v_A}H_{AA}\ket{v_A} +
\bra{v_B}H_{BB}\ket{v_B} -
\bra{v_B}H_{BA}\ket{v_A} -
\bra{v_A}H_{AB}\ket{v_B} \\
    &= \lambda_1 - 4 \bra{v_B}H_{BA}\ket{v_A}.
\end{align*}
So if we can show that 
$ \bra{v_B}H_{BA}\ket{v_A}$
tends to zero we are done.
This term involves the 
dynamical coupling between the
groundstate wavefunction along
the cut between $A$ and $B$.
To succeed we must find such a cut where
the wavefunction is small. In general
this appears to be quite difficult,
even though in the models we are considering
numerics show that not only is the
wavefunction small away from potential wells
but it is exponentially small.


\subsection{The cut and symmetry}\label{Sec72}

We now study 
the cut associated to the second eigenvector of a 
weakly self-dual gauge Hamiltonian $H,$
and relate this to the stabilizers of the code.
The key realization is that $\Gamma_{t_X}$ is like
the double well potential above,
but now we have $2^{m_X}$ such wells,
that is, one for every $s_X\in \Span{S_X}.$
This is clear from examining the basis vectors for $\Gamma_{t_X}.$
These are 
$$
    \ket{v S_X + u R_X + t_X}, \smbox{where} v\in \Field_{m_X}, u\in\Frd
$$
and those that satisfy the most $G_Z$ terms are
precisely those with $u=0.$

We already know this is either the second eigenvector of $H_{0,0}$
or otherwise the first eigenvector of $H_{t_X,0}$ for some $t_X \ne 0.$
To relate this to the Perron-Frobenius theory we note the 
decomposition from Lemma \ref{lemma26}:
$$
    \Gamma_{t_X} = \bigoplus_{t_Z\in\Span{T_Z}} H_{t_X, t_Z}.
$$
This gives the spectral decomposition of each graph $\Gamma_{t_X}$ 
in terms of ``momenta'' $t_Z.$

We focus on $\Gamma_0.$
This must contain the second eigenvector of $H$ by weak self-duality of the code.
$X$-type stabilizers $s_X\in S_X$ act on the $0,t_Z$ irreps in $\Gamma_0$
by $\pm 1$ according to the commutator $[[s_X, t_Z]].$
Suppose the second eigenvector of $H$ lives in
$H_{0,t_Z}$ for $t_Z\ne 0$. 
Let $s_X\in S_X$ with $[[s_X, t_Z]]=-1.$
Then we must have an odd number of Cheeger cuts 
on every $\Gamma_0$ path between $\ket{v}$ and $s_X\ket{v}$ for all basis
vectors $\ket{v},$ that is, $v\in\Span{S_X}\oplus\Span{R_X}.$

In a similar vein, if the second eigenvector of $H$ lives in $H_{0,0}$
then we must have an even number of Cheeger cuts 
on every $\Gamma_0$ path between $\ket{v}$ and $s_X\ket{v}$ for all stabilizers $s_X\in S_X$
and basis vectors $\ket{v}.$

In summary, the idea is that large stabilizers lead to
widely separated well potentials and hence gapless behaviour,
while stabilizers of bounded weight force the cuts to
appear close to the wells and hence maintain a gap.
Even though numerics show the wavefunction becoming 
exponentially small away from well potentials,
it is also exponentially wide.
So making these arguments rigorous appears to be difficult.

The following fact would appear to be true under certain conditions,
but is not at all true for example when $T$ is trivial:
\begin{framed}
\noindent{\bf Proto-fact:}
For a sufficiently ``well-behaved''
weakly self-dual
gauge code Hamiltonian $H$
\begin{align*}
\lambda_2(\Ham) 
    &= \min_{t_X\ne 0} \lambda_1(\Ham_{t_X,0})\\
    &= \min_{t_Z\ne 0} \lambda_1(\Ham_{0,t_Z}).
\end{align*}
\end{framed}
Indeed, contrary to this proto-fact
we suspect that $H_{0,0}$ will not be gapped in
the generic case. 
Numerics suggest that
there is no lower bound on the gap of 
randomly constructed stabilizer-less gauge code Hamiltonians.
Perhaps double well behaviour can still be imitated even without
stabilizers: merely having a large region of almost-stabilizer
behaviour (large shallow well) could be enough to send the gap to zero.


\subsection{Cheeger inequalities}\label{Sec73}

We saw above how the Cheeger cut gives a variational ansatz
for building a second eigenvector to the Hamiltonian and hence
an upper bound on the gap.
In this section we show how the Cheeger cut also 
yields a lower bound on the gap.

In \cite{Friedland2002}, they derive the following Cheeger inequality
by considering bi-partitions of the graph. We will do the
same, but using matrix block notation.

Let $v_2$ be a second eigenvector, $ \Ham v_2 = \lambda_2 v_2 $ 
and $||v_2||=1$.
We bi-partition the space 
so that $v_2$ has (vector) blocks:
$$
v_2 = \left( \begin{array}{l}
x\\
y\end{array} \right)\quad
$$
with $x\ge 0$ and $y\le 0,$ entry-wise.
Let the blocks of $\Ham$ under the same partition be:
$$
\Ham = \left( \begin{array}{ll}
A&C\\
C^\top&B\end{array} \right).\quad
$$
If we denote $\lambda_1(A)$ as the top eigenvalue of $A$ and
$\lambda_1(B)$ as the top eigenvalue of $B$,
then
\begin{align*}
\lambda_2 = v_2^\top \Ham v_2 &= x^\top A x + 2 x^\top C y + y^\top B y \\
        &\le x^\top A x + y^\top B y\ \le\ ||x||^2 \lambda_1(A) + ||y||^2 \lambda_1(B) \\
        &\le \mbox{min}(\lambda_1(A), \lambda_1(B))\ \le\ \lambda_1.
\end{align*}

Defining the following constant as a maximization over
all bi-partitions of $\Ham:$
$$
    \nu(\Ham) := \max_{A, B}\ \mbox{min}(\lambda_1(A), \lambda_1(B))
$$
the above calculation shows that
$$
    \lambda_2 \le \nu(\Ham) \le \lambda_1.
$$


\section{Discussion}\label{Sec8}

The goal of this paper is to understand the spectra of
certain frustrated qubit Hamiltonians. Of particular interest is
to understand how 
the gap between the first two eigenvalues behaves as we 
increase the number of qubits.
For a system to maintain a topologically ordered groundstate
we would expect this gap to be bounded away from zero, or gapped.

The numerical results obtained show some evidence for
this gapped behaviour for the 3D gauge color code Hamiltonian.
To get these numerics we rely on several key results.
The first is to decompose the Hamiltonian into a direct sum
of operators labelled by stabilizer eigenvalues.
This is essentially group representation theory as applied to
gauge codes.
The second result was obtained using Perron-Frobenius theory, as the
Hamiltonians of interest are stoquastic.
This theory shows where the first and second eigenvalue
is to be found in the block decomposition of the Hamiltonian,
which leads to a polynomial reduction in the numeric workload.
The third key result is the decomposition of
each Hamiltonian block into mutually commuting ideals.
In the case of the 3D gauge color code Hamiltonian, this
decomposition yields an exponential reduction in 
the numeric workload, by dividing the number of qubits by six.

Finally, the connection between stabilizer size and the gap is further 
investigated via the idea of the Cheeger cut. 
This builds on the Perron-Frobenius theory results.
If the groundstate wavefunction is sufficiently concentrated
into potential wells, then we can construct an excited state
that has eigenvalue close to the ground eigenvalue.
And the distance between potential wells is controlled by
the size of stabilizers.
The conjecture we would like to make would state 
how this spectral gap is controlled by the size of the frustrated
stabilizers: large stabilizers lead to gappless behaviour while
small stabilizers maintain a gap.
However, it is not clear how to formulate this conjecture, and
even less clear how to prove it. 
More numerics need to be performed, with the specific
goal of understanding the shape of the groundspace wavefunction,
and how this relates to the geometry of the underlying code.

\bibliography{refs3}{}
\bibliographystyle{abbrv}

\end{document}